\newif\iffigs\figstrue
   \documentclass[12pt]{article}
\iffigs
   \input{epsf}
\else
\fi
\newcommand{\eqn}[1]{(\ref{#1})}

\newsavebox{\uuunit}
\sbox{\uuunit}
    {\setlength{\unitlength}{0.825em}
     \begin{picture}(0.6,0.7)
        \thinlines
        \put(0,0){\line(1,0){0.5}}
        \put(0.15,0){\line(0,1){0.7}}
        \put(0.35,0){\line(0,1){0.8}}
       \multiput(0.3,0.8)(-0.04,-0.02){12}{\rule{0.5pt}{0.5pt}}
     \end {picture}}

\def\IP{\relax{\rm I\kern-.18em P}}

\begin{document}
%
%%%%%%%%%%%%%%%%%%%%%%%%%%%%%%%%%%%%%%%%%%%%%%%%%%%%%%%%%%
% Definizioni Varie %%%%%%%%%%%%%%%%%%%%%%%%%%%%%%%%%%%%%%
%%%%%%%%%%%%%%%%%%%%%%%%%%%%%%%%%%%%%%%%%%%%%%%%%%%%%%%%%%
%   *****  Some definitions  *****
\font\cmss=cmss10 \font\cmsss=cmss10 at 7pt
\def\twomat#1#2#3#4{\left(\matrix{#1 & #2 \cr #3 & #4}\right)}
\def\inbar{\vrule height1.5ex width.4pt depth0pt}
\def\IC{\relax\,\hbox{$\inbar\kern-.3em{\rm C}$}}
\def\IG{\relax\,\hbox{$\inbar\kern-.3em{\rm G}$}}
\def\IB{\relax{\rm I\kern-.18em B}}
\def\ID{\relax{\rm I\kern-.18em D}}
\def\IL{\relax{\rm I\kern-.18em L}}
\def\IF{\relax{\rm I\kern-.18em F}}
\def\IH{\relax{\rm I\kern-.18em H}}
\def\II{\relax{\rm I\kern-.17em I}}
\def\IN{\relax{\rm I\kern-.18em N}}
\def\IP{\relax{\rm I\kern-.18em P}}
\def\IQ{\relax\,\hbox{$\inbar\kern-.3em{\rm Q}$}}
\def\bfzero{\relax\,\hbox{$\inbar\kern-.3em{\rm 0}$}}
\def\IR{\relax{\rm I\kern-.18em R}}
\def\ZZ{\relax\ifmmode\mathchoice
{\hbox{\cmss Z\kern-.4em Z}}{\hbox{\cmss Z\kern-.4em Z}}
{\lower.9pt\hbox{\cmsss Z\kern-.4em Z}}
{\lower1.2pt\hbox{\cmsss Z\kern-.4em Z}}\else{\cmss Z\kern-.4em
Z}\fi}
\def\bfone{\relax{\rm 1\kern-.35em 1}}
\def\dop{{\rm d}\hskip -1pt}
\def\real{{\rm Re}\hskip 1pt}
\def\trace{{\rm Tr}\hskip 1pt}
\def\ii{{\rm i}}
\def\diag{{\rm diag}}
\def\sch#1#2{\{#1;#2\}}
%%%%%%%%%%%%%%%%%%%%%%%%%%%%%%%%%%%%%%%%%%%%%%%%%%%%%%%%%%%%
% new commands  %%%%%%%%%%%%%%%%%%%%%%%%%%%%%%%%%%%%%%%%%%%%%
% ANNAMACRO.TEX %%%%%%%%%%%%%%%%%%%%%%%%%%%%%%%%%%%%%%%%%%%%%
%%%%%%%%%%%%%%%%%%%%%%%%%%%%%%%%%%%%%%%%%%%%%%%%%%%%%%%%%%%%%
\def\bfone{\relax{\rm 1\kern-.35em 1}}
\font\cmss=cmss10 \font\cmsss=cmss10 at 7pt
\def\a{\alpha} \def\b{\beta} \def\d{\delta}
\def\e{\epsilon} \def\c{\gamma}
\def\G{\Gamma} \def\l{\lambda}
\def\L{\Lambda} \def\s{\sigma}
\def\cA{{\cal A}} \def\cB{{\cal B}}
\def\cC{{\cal C}} \def\cD{{\cal D}}
\def\cF{{\cal F}} \def\cG{{\cal G}}
\def\cH{{\cal H}} \def\cI{{\cal I}}
\def\cJ{{\cal J}} \def\cK{{\cal K}}
\def\cL{{\cal L}} \def\cM{{\cal M}}
\def\cN{{\cal N}} \def\cO{{\cal O}}
\def\cP{{\cal P}} \def\cQ{{\cal Q}}
\def\cR{{\cal R}} \def\cV{{\cal V}}\def\cW{{\cal W}}
\newcommand{\be}{\begin{equation}}
\newcommand{\ee}{\end{equation}}
\newcommand{\bea}{\begin{eqnarray}}
\newcommand{\eea}{\end{eqnarray}}
\let\la=\label \let\ci=\cite \let\re=\ref
%
%
%%%%%%%%%%%%%%%%%%%%%%%%%%%%%%%%%%%%%%%%%%%%%%%%%%%%%%%%%%%%%%
%%%%% misc macros %%%%%
%%%%%%%%%%%%%%%%%%%%%%%%%%%%%%%%%%%%%%%%%%%%%%%%%%%%%%%%%%%%%%
%
\def\crr{\crcr\noalign{\vskip {8.3333pt}}}
\def\tilde{\widetilde}
\def\bar{\overline}
\def\us#1{\underline{#1}}
\let\shat=\hat
\def\hat{\widehat}
\def\hyp{\vrule height 2.3pt width 2.5pt depth -1.5pt}
\def\square{\mbox{.08}{.08}}
\def\Coeff#1#2{{#1\over #2}}
\def\Coe#1.#2.{{#1\over #2}}
\def\coeff#1#2{\relax{\textstyle {#1 \over #2}}\displaystyle}
\def\coe#1.#2.{\relax{\textstyle {#1 \over #2}}\displaystyle}
\def\half{{1 \over 2}}
\def\shalf{\relax{\textstyle {1 \over 2}}\displaystyle}
\def\dag#1{#1\!\!\!/\,\,\,}
\def\to{\rightarrow}
\def\notin{\hbox{{$\in$}\kern-.51em\hbox{/}}}
\def\shdot{\!\cdot\!}
\def\ket#1{\,\big|\,#1\,\big>\,}
\def\bra#1{\,\big<\,#1\,\big|\,}
\def\equaltop#1{\mathrel{\mathop=^{#1}}}
\def\Trbel#1{\mathop{{\rm Tr}}_{#1}}
\def\inserteq#1{\noalign{\vskip-.2truecm\hbox{#1\hfil}
\vskip-.2cm}}
\def\attac#1{\Bigl\vert
{\phantom{X}\atop{{\rm\scriptstyle #1}}\phantom{X}}}
\def\exx#1{e^{{\displaystyle #1}}}
\def\del{\partial}
\def\delbar{\bar\partial}
\def\nex#1{$N\!=\!#1$}
\def\dex#1{$d\!=\!#1$}
\def\cex#1{$c\!=\!#1$}
\def\eg{{\it e.g.}} \def\ie{{\it i.e.}}
%\catcode`\@=12
%%%%%%%%%%%%%%%%%%%%%%%%%%%%%%%%%%%%%%%%%%%%%%%%%%%%%%%%%%%%%%
%%%%%%%%%%%%%%%%%%%%%%%%%%%%%%%%%%%%%%%%%%%%%%%%%%%%%%%%%%%%
%\draft
%%%%%%%%%%%% macros and references %%%%%%%%%%%%%%%%%%%%%%%%%
\def\IE{\relax{{\rm I\kern-.18em E}}}
\def\cE{{\cal E}}
\def\rt{{\cR^{(3)}}}
\def\IGam{\relax{{\rm I}\kern-.18em \Gamma}}
\def\IGa{\IA}
\def\ii{{\rm i}}
%%%%%%%%%%%%%%%%%%%%%%%%%%%%%%%%%%%%%%%%%%%%%%
%%%%%%%%%%%%%%%%%%%%%%%%%%%%%%%%%%%%%%%%%%%%%
%%%%%%%%%%%%%%%%%%%%%%%%%%%%%%%%%%%%%%%%%%%%%%%%%%%%%%%%%%%
\begin{titlepage}
\begin{center}
{\LARGE U--Duality, Solvable Lie Algebras \\
\vskip 1.5mm
and Extremal Black--Holes$^\star$\\
}
\vfill
{\it Talk given at the III National Meeting \\
of the Italian Society for General Relativity (SIGRAV)\\
on the occasion of Prof. Bruno Bertotti's 65th birthday\\
Rome September 1996}\\
\vfill
{\large {\bf Pietro Fr\'e}} \\
\vfill
{\small
 Dipartimento di Fisica Teorica, Universit\`a di Torino, via P. Giuria 1,
I-10125 Torino, \\
 Istituto Nazionale di Fisica Nucleare (INFN) - Sezione di Torino, Italy \\
}
\end{center}
\vfill
\begin{center}
{\bf Abstract}
\end{center}
{\small In this lecture I review recent results on the use of Solvable Lie Algebras
as an  efficient description of the scalar field sector of supergravities
in relation with their non perturbative structure encoded in the U--duality group.
I also review recent results on the construction of BPS saturated states
as solution of the first differential equations following from imposing
preservation of a fraction of the original supersymmetries.
In particular I discuss $N=2$ extremal
black--holes that are approximated by a Bertotti--Robinson metric near their horizon.
The extension of this construction to maximally extended supergravities in all
dimensions $4 \le D \le 11$ is work in progress where the use of
the Solvable Lie algebra approach promises to be of decisive
usefulness. }
\vspace{2mm} \vfill \hrule width 3.cm
{\footnotesize
 $^*$ Supported in part by   EEC  under TMR contract
 ERBFMRX-CT96-0045. }
\end{titlepage}
%%%%%%%%%%%%%%%%%%%%%%%%%%%%%%%%%%%%%%%%%%%%%%%%%%%%%%%

\section{Introduction}
Supersymmetry has been the most powerful tool to advance
our understanding of quantum field theory and, in its $N=1,D=4$ spontaneously broken
form, it might also be experimentally observable. The locally $N$--extended
supersymmetric field theories, namely $N$--extended supergravities,
are interpreted as the low energy effective actions of various superstring
theories in diverse dimensions. However, since the duality revolution of
two years ago \cite{sumschwarz}, we know that all superstring models
(with their related effective actions) are just different corners of
a single non--perturbative quantum theory that includes, besides
strings also other $p$--brane excitations. Indeed in dimensions $D
\ne 4$ the duality rotations \cite{mylec} from
electrically to magnetically charged particles (= $0$--branes) generalize  to
transformations exchanging the perturbative elementary $p$--brane excitations
of a theory with the non--perturbative solitonic
$D-p-4$--brane excitations of the same theory. The mass per unit world--volume of
these objects is lower bounded by the value of the topological central
charge according to a generalization of the classical Bogomolny bound
on the monopole mass. In the recent literature,
the states saturating this lower bound are named  BPS saturated states
\cite{BPS} and play a prominent role in establishing the exact duality
symmetry of the quantum theory since they are the lowest lying stable
states of the non perturbative spectrum.
\par
The key instrument to connect the perturbative sector
of one version  of string theory to the perturbative sector
of another version reinterpreting the weak coupling regime
of the latter as the strong coupling regime of the former
comes from a well established feature of supergravity theories
discovered in the earliest stages of their development, namely the
so called {\it hidden symmetries}.
%%%%%%%%%%%%%%%%
Indeed the hidden non compact symmetries  of extended supergravities \cite{cre}
already discovered in the late seventies and beginning of the eighties
have recently played a major role in unravelling
some non perturbative properties
of string theories such as various types of dualities occurring in different dimensions and in
certain regions of the moduli spaces \cite{schw}.
\par
In particular their discrete remnants have been of
crucial importance to discuss, in a model independent way, some physical properties
such as the spectrum of BPS states \cite{sesch}, \cite{hamo}, \cite{huto2}
and entropy formulas for the already mentioned extreme black--holes
\cite{feka}\cite{malda}.
\par
It is common wisdom that such U--dualities should play an important role in the understanding
of other phenomena such as the mechanism for supersymmetry breaking, which may be
due to some non perturbative physics
 \cite{postr},\cite{wi},\cite{klmvw}.
\par
Recently, in collaboration with D'Auria, Ferrara, Andrianopoli and Trigiante,
the present author  has analyzed some properties of U--duality symmetries
in any dimensions in the context of solvable Lie algebras \cite{solvab1,solvab2}.
\par
In string theories or M--theory compactified to lower dimensions \cite{witten},
preserving $N>2$ supersymmetries, the U--duality group  is generically
an infinite dimensional discrete subgroup $U(\ZZ) \subset U$,
where $U$ is related to the non--compact symmetries of the low energy effective
supergravity theory \cite{huto}.
\par
The solvable Lie algebra $G_S=Solv(U/H)$ with the property $\exp [G_S] = U/H$,
where $U/H$ is (locally) the scalar manifold of the theory, associates
group generators to each scalar, so that one can speak of NS and R--R generators.
\par
Translational symmetries of NS and/or R--R fields are associated with the maximal
abelian nilpotent ideal of $\cA \subset G_S$, with a series of implications.
\par
The advantage of introducing such notion is twofold:
besides that  of associating generators with  scalar fields when decomposing
the U--duality group with respect to perturbative and non perturbative
symmetries of string theories, such as T and S--duality in type IIA
 or $SL(2,\IR)$
duality in type IIB, one may unreveal connections between different theories
and have an understanding of N--S and R--R generators at the group--theoretical
level, which may hold beyond a particular perturbative framework.
Furthermore, the identification $U/H \sim \exp[G_S]$ of the scalar coset manifold
with the group manifold of a normed solvable Lie algebra allows the description
of the local differential geometry of $U/H$ in purely algebraic terms.
Since the effective low energy supergravity lagrangian is entirely encoded in terms of this
local differential geometry, this fact has obvious distinctive advantages.
\par
In {solvab1,solba2}, we have derived a certain number of relations among
 solvable Lie algebras which explain some of the results obtained by some of
us in a previous work.
\par
In particular, we have shown that the Peccei--Quinn (translational) symmetries
of $U_{D}/H_{D}$
in $D=10-r$ dimensions are classified  by $U_{D+1}$, while their
 NS and R--R content are classified by $O(r-1,r-1)$.
\par
For $D>3$ this content corresponds to the number of vector fields in the $D+1$
theory, at least for maximal supergravities.
\par
An explicit expression for these generators was given and an interpretation
in terms of branes was also provided. This shows the close
relationship between the existence and structure of extreme
black-hole or black--brane solutions of effective supergravity
theories and the conjectured U--duality symmetry characterizing the
non--perturbative string theory.
\par
The glue connecting all these issues
is supersymmetry.
%%%%%%%%%%%%%%%%%%%%%%%%%
The BPS saturated states are characterized by the
fact that they preserve, in modern parlance, $1/2$ (or $1/4$, or $1/8$) of the original
supersymmetries. What this actually means is that there is a suitable
projection operator $\IP^2_{BPS} =\IP_{BPS}$ acting on the supersymmetry charge
$Q_{SUSY}$, such that:
\begin{equation}
 \left(\IP_{BPS} \,Q_{SUSY} \right) \, \vert \, \mbox{BPS state} \, >
 \,=  \, 0
 \label{bstato}
\end{equation}
Since the supersymmetry transformation rules of any supersymmetric
field theory are linear in the first derivatives of the fields
eq.\eqn{bstato} is actually a {\it system of first order differential
equations} such that any solution of \eqn{bstato} is also a solution
of the {\it second order field equations} derived from the action
but not viceversa. Hence the case of BPS states, that includes
extremal monopole and black--hole configurations, is an instance of
the relation existing between {\it supersymmetry and first--order square
roots of the classical field equations}. Another important example of this
relation is provided by the {\it the topological twist}
\cite{topftwist_1,topftwist_2,topf4d_8} of
supersymmetric theories to {\it topological field theories}
\cite{wittft}. What happens here is that, after Wick rotation to the
Euclidean region, there is another projection operator $\IP_{BRST}^2=
\IP_{BRST}$ acting on the supersymmetry charge $Q_{SUSY}$, such that:
\begin{equation}
 \left(\IP_{BRST} \,Q_{SUSY} \right) \, \vert \, \mbox{Instanton} \, >
 \,=  \, 0
 \label{inststato}
\end{equation}
In the case of topological field theories the projected supersymmetry
charge is interpreted as the $BRST$--charge $Q_{BRST}$ associated with the
topological symmetry and the generalized instanton configurations
satisfying eq.\eqn{inststato}, being in the kernel of $Q_{BRST}$, are
representatives of the cohomology classes of {\it physical states}.
For the same reason as above \eqn{inststato} are {\it first order
differential equations}. \par
%%%%%%%%%%%%%%%%%%%%%%%
In this lecture I will illustrate the relation between supersymmetry
and the  first order differential equations for BPS states
with examples taken from both rigid and local
$N=2$ theories in $D=4$. Here the recently obtained fully general
form of $N=2$ SUGRA and $N=2$ SYM \cite{jgpnoi} allows to match the
structure of Special K\"ahler geometry
\cite{specspec2,skgsugra_4,skgsugra_1} with the structure of the
first order differential equations. In particular a vast number of
results were recently obtained for the case of $N=2$ extremal
black--holes \cite{cardoso,kalvanp1,ferkal2}, \cite{strom3,ferkal4,kalmany}.
The main idea will be reviewed in section ~\ref{BPSlocala}. I have
reviewed this material  in a slightly different perspective in
another talk \cite{santama}, where my emphasis was more focused on
a comparison between the first order equations emerging in the BPS
problem and those emerging in the generalized instanton case.
\par
In the present talk, as an introduction to work in progress
 on the application
of solvable Lie algebras to the first order equations for
$BPS$--saturated states, my emphasis will be shifted
to the solvable Lie algebra aspects.
\par
In {solvab1,solvab2}
we compared different decompositions of solvable Lie algebras
in IIA, IIB and M--theory in toroidal
compactifications which preserve maximal supersymmetry (i.e. 32 supercharges).
\par
While in Type IIA the relevant decomposition is with respect to the S--T
duality group, in IIB theory we decomposed the U--duality group with respect
to $SL(2,\IR)\times GL(r,\IR)$ and in M--theory with respect to $GL(r+1, \IR)$.
\par
Comparison of these decompositions shows some of the non-perturbative
relations existing among these theories, such as the interpretation of
$SL(2,\ZZ)$ as the group acting on the complex structure of a two-dimensional
 torus \cite{vasch}.
\par
Solvable Lie algebras play also an important role in the gauging of isometries
while preserving vanishing cosmological constant or partially breaking
some of the supersymmetries. Indeed, this was used in the literature
 \cite{fegipo} in the
 context of $N=2$ supergravity spontaneously broken to $N=1$ and may be used
 in a more general framework. This study is relevant in view of possible applications
 in string effective field theories, where field-strength condensation
may give rise to the gauging of abelian isometries \cite{postr} generating
a flat potential that spontaneously breaks supersymmetry.
\par
Summarizing the present discussion it is clear that the
subject of extremal black--brane configurations in supersymmetric
field theories and the subject of the Solvable Lie algebra description
of U--dualities are intimately and surprisingly related. This is the
main motivation for discussing them at the same time in this talk.
It is also very remarkable that in the $4$--dimensional case the
universality class of the metrics that the  realize the non--perturbative
BPS spectrum implied by U--duality is given by the
time--honoured {\it Bertotti Robinson metric}. Indeed this latter
characterizes the universal behaviour of all spherically symmetric
BPS black holes near their horizon. It is with my utmost pleasure
that this talk offers me the opportunity to express Prof. Bruno Bertotti
my admiration for his scholarship and scientific achievements and
also my most friendly and sincere best wishes for a happy birthday.

%%%%%%%%%%%%%%%%%%%%%%%%%%%%%%%%%%%%%%%%%%%%%%%%%%%%%%%%%%%%%%%%%%%%%%%
\vskip 0.2cm
\section{Central Charges}
\label{centcharge}
Let us consider the $D=4$ supersymmetry algebra with an even number
$N=2\nu$ of supersymmetry charges. It can be written in the following
form:
\begin{eqnarray}
&\left\{ {\bar Q}_{Ai \vert \alpha }\, , \,{\bar Q}_{Bj \vert \beta}
\right\}\, = \,  {\rm i} \left( C \, \gamma^a \right)_{\alpha \beta} \,
P_a \, \delta_{AB} \, \delta_{ij} \, - \, C_{\alpha \beta} \,
\epsilon_{AB} \, \times \, \ZZ_{ij}& \nonumber\\
&\left( A,B = 1,2 \qquad ; \qquad i,j=1,\dots, \nu \right)&
\label{susyeven}
\end{eqnarray}
where the SUSY charges ${\bar Q}_{Ai}\equiv Q_{Ai}^\dagger \gamma_0=
Q^T_{Ai} \, C$ are Majorana spinors, $C$ is the charge conjugation
matrix, $P_a$ is the 4--momentum operator, $\epsilon_{AB}$ is the
two--dimensional Levi Civita symbol and the symmetric tensor
$\ZZ_{ij}=\ZZ_{ji}$ is the central charge operator. It
can always be diagonalized $\ZZ_{ij}=\delta_{ij} \, Z_j$ and its $\nu$ eigenvalues
$Z_j$ are the central charges.
\par
The Bogomolny bound on the mass of a generalized monopole state:
\begin{equation}
M \, \ge \, \vert \, Z_i \vert \qquad \forall Z_i \, , \,
i=1,\dots,\nu
\label{bogobound}
\end{equation}
is an elementary consequence of the supersymmetry algebra and of the
identification between {\it central charges} and {\it topological
charges}. To see this it is convenient to introduce the following
reduced supercharges:
\begin{equation}
{\bar S}^{\pm}_{Ai \vert \alpha }=\frac{1}{2} \,
\left( {\bar Q}_{Ai}\pm \mbox{i} \, \epsilon_{AB} \,  {\bar Q}_{Bi}\,
\right)_\alpha
\label{redchar}
\end{equation}
They can be regarded as the result of applying
a projection operator to the supersymmetry
charges:
\begin{eqnarray}
{\bar S}^{\pm}_{Ai} &=& {\bar Q}_{Bi} \, \IP^\pm_{BA} \nonumber\\
 \IP^\pm_{BA}&=&\frac{1}{2}\, \left({\bf 1}\delta_{BA} \pm \mbox{i} \epsilon_{BA}
 \gamma_0 \right)
 \label{projop}
\end{eqnarray}
Combining eq.\eqn{susyeven} with the definition \eqn{redchar} and
choosing the rest frame where the four momentum is $P_a$ =$(M,0,0,0)$, we
obtain the algebra:
\begin{equation}
\left\{ {\bar S}^{\pm}_{Ai}  \, , \, {\bar S}^{\pm}_{Bj} \right\} =
\pm \epsilon_{AC}\, C \, \IP^\pm_{CB} \, \left( M \mp Z_i \right)\,
\delta_{ij}
\label{salgeb}
\end{equation}
By positivity of the operator $\left\{ {S}^{\pm}_{Ai}  \, , \, {\bar S}^{\pm}_{Bj} \right\} $
it follows that on a generic state the Bogomolny bound \eqn{bogobound} is
fulfilled. Furthermore it also follows that the states which saturate
the bounds:
\begin{equation}
\left( M\pm Z_i \right) \, \vert \mbox{BPS state,} i\rangle = 0
\label{bpstate1}
\end{equation}
are those which are annihilated by the corresponding reduced supercharges:
\begin{equation}
{\bar S}^{\pm}_{Ai}   \, \vert \mbox{BPS state,} i\rangle = 0
\label{susinvbps}
\end{equation}
%%%%%%%%%%%%%%%%%%%%%%%%%%%%%%%%%%%%%%%%%%%%%%%%%%%
% Cassandra %%%%%%%%%%%%%%%%%%%%%%%%%%%%%%%%%%%%%%%
%%%%%%%%%%%%%%%%%%%%%%%%%%%%%%%%%%%%%%%%%%%%%%%%%%%
\section{The solvable Lie algebra:  NS and RR scalar fields}
It has been known for many years \cite{sase} that the scalar field manifold of
both pure and matter coupled
$N>2$ extended supergravities in $D=10-r$ ($r=6,5,4,3,2,1$) is a non compact
homogenous symmetric manifold $U_{(D,N)} /H_{(D,N)}$, where $U_{(D,N)}$ (depending
on the space--time dimensions and on the number of supersymmetries) is a non compact
Lie group and $H_{(D,N)}\subset U_{(D,N)}$ is a maximal compact subgroup.
For instance in the physical $D=4$ case the situation is summarized
in table \ref{topotable}.
%%%%%%%%%%%%%%%%%%%%%%%%%%%%%%%%%%%%%%%%
% plutarco %%%%%%%%%%%%%%%%%%%%%%%%%%%%%
%%%%%%%%%%%%%%%%%%%%%%%%%%%%%%%%%%%%%%%%
{\footnotesize
\begin{table*}
\begin{center}
\caption{\sl Scalar Manifolds of Extended Supergravities}
\label{topotable}
\begin{tabular}{|c||c|c|c||c|c||c||c| }
\hline
\hline
~ & $\#$ scal. & $\#$ scal. & $\#$ scal. & $\#$ vect. &
 $\#$ vect. &~ & $~ $ \\
N & in & in & in & in  &
 in  &$\Gamma_{cont}$ & ${\cal M}_{scalar}$   \\
 ~ & scal.m. & vec. m. & grav. m. & vec. m. & grav. m. & ~ &~
\\
\hline
\hline
~    &~    &~   &~   &~  &~  & ~ & ~ \\
$1$  & 2 m &~   & ~  & n &~  &  ${\cal I}$  & ~   \\
~    &~    &~   &~   &~  &~  &  $\subset Sp(2n,\IR)$ & K\"ahler \\
~    &~    &~   &~   &~  &~  & ~ & ~ \\
\hline
~    &~    &~   &~   &~  &~  & ~ & ~ \\
$2$  & 4 m & 2 n& ~  & n & 1 &  ${\cal I}$ & Quaternionic $\otimes$
\\
~    &~    &~   &~   &~  &~  &  $\subset Sp(2n+2,\IR)$ & Special K\"ahler \\
~    &~    &~   &~   &~  &~  & ~ & ~ \\
\hline
~    &~    &~   &~   &~  &~  & ~ & ~ \\
$3$  & ~   & 6 n& ~  & n & 3 &  $SU(3,n)$ &~  \\
~    &~    &~   &~   &~  &~  & $\subset Sp(2n+6,\IR)$ & $\frac{SU(3,n)}
{S(U(3)\times U(n))}$ \\
~    &~    & ~  &~   &~  &~  & ~ & ~ \\
\hline
~    &~    &~   &~   &~  &~  & ~ & ~ \\
$4$  & ~   & 6 n& 2  & n & 6 &  $SU(1,1)\otimes SO(6,n)$ &
$\frac{SU(1,1)}{U(1)} \otimes $ \\
~    &~    &~   &~   &~  &~  & $\subset Sp(2n+12,\IR)$ &
$\frac{SO(6,n)}{SO(6)\times SO(n)}$ \\
~    &~    &~   &~   &~  &~  & ~ & ~ \\
\hline
~    &~    &~   &~   &~  &~  & ~ & ~ \\
$5$  & ~   & ~  & 10 & ~ & 10 & $SU(1,5)$ & ~  \\
~    &~    &~   &~   &~  &~  & $\subset Sp(20,\IR)$ & $\frac{SU(1,5)}
{S(U(1)\times U(5))}$ \\
~    &~    &~   &~   &~  &~  & ~ & ~ \\
\hline
~    &~    &~   &~   &~  &~  & ~ & ~ \\
$6$  & ~   & ~  & 30 & ~ & 16 & $SO^\star(12)$ & ~ \\
~    &~    &~   &~   &~  &~  & $\subset Sp(32,\IR)$ &
$\frac{SO^\star(12)}{U(1)\times SU(6)}$ \\
~    &~    &~   &~   &~  &~  & ~ & ~ \\
\hline
~    &~    &~   &~   &~  &~  & ~ & ~ \\
$7,8$& ~   & ~  & 70 & ~ & 56 & $E_{7(-7)}$  & ~ \\
~    &~    &~   &~   &~  &~  & $\subset Sp(128,\IR)$ &
$\frac{ E_{7(-7)} }{SU(8)}$ \\
~    &~    &~   &~   &~  &~  & ~ & ~ \\
\hline
 \hline
\end{tabular}
\end{center}
\end{table*}
}
Furthermore, the structure of the supergravity lagrangian is completely encoded in the
local differential geometry of
$U_{(D,N)}/H_{(D,N)}$.
\par
The recent exciting developments on the non--perturbative structure
of string theory have started from the conjecture \cite{huto} that
an appropriate restriction to integers $U_{(D,N)}(\ZZ)$ of
the Lie group $U_{(D,N)}$
is an exact non perturbative symmetry of string theory. Eventually it
permutes the elementary, electric states of the perturbative string
spectrum with the non perturbative BPS saturated states like the
Black Holes discussed in later sections of this talk. This U--duality
unifies S--duality (strong--weak duality) with T--duality
(large--small radius duality).
\par
As discussed in \cite{solvab1,solvab2}, utilizing a well established mathematical
framework \cite{helgason}, in all these cases the scalar coset manifold $U/H$ can be
identified with the group manifold of a normed solvable Lie algebra:
\begin{equation}
  U/H \sim \exp[{Solv}]
\end{equation}
\par
The representation of the supergravity scalar manifold ${\cal M}_{scalar}= U/H$
as the group manifold associated with a {\it  normed solvable Lie algebra}
introduces a one--to--one correspondence between the scalar fields $\phi^I$ of
supergravity and the generators $T_I$ of the solvable Lie algebra $Solv\, (U/H)$.
Indeed the coset representative $L(U/H)$ of the homogeneous space $U/H$ is
identified with:
\begin{equation}
L(\phi) \, =\, \exp [ \phi^I \, T_I ]
\label{cosrep1}
\end{equation}
where $\{ T_I \}$ is a basis of $Solv\, (U/H)$.
\par
As a consequence of this fact the tangent bundle to the scalar manifold $T{\cal M}_{scalar}$
is identified with the solvable Lie algebra:
\begin{equation}
T{\cal M}_{scalar} \, \sim \,Solv \, (U/H)
\label{cosrep2}
\end{equation}
and any algebraic property of the solvable algebra has a corresponding physical interpretation in
terms of string theory massless field modes.
\par
Furthermore, the local differential geometry of the scalar manifold is described
 in terms of the solvable Lie algebra structure.
Given the euclidean scalar product on $Solv$:
\begin{eqnarray}
  <\, , \, > &:& Solv \otimes Solv \rightarrow \IR
\label{solv1}\\
<X,Y> &=& <Y,X>\label{solv2}
\end{eqnarray}
the covariant derivative with respect to the Levi Civita connection is given by
the Nomizu operator \cite{alex}:
\begin{equation}
\forall X \in Solv : \IL_X : Solv \to Solv
\end{equation}
\begin{eqnarray}
  \forall X,Y,Z \in Solv & : &2 <Z,\IL_X Y> \nonumber\\
&=& <Z,[X,Y]> - <X,[Y,Z]> - <Y,[X,Z]>
\label{nomizu}
\end{eqnarray}
and the Riemann curvature 2--form is given by the commutator of two Nomizu
operators:
\begin{equation}
 <W,\{[\IL_X,\IL_Y]-\IL_{[X,Y]}\}Z> = R^W_{\ Z}(X,Y)
\label{nomizu2}
\end{equation}
In the case of maximally extended supergravities in $D=10-r$ dimensions the scalar
manifold has a universal structure:
\begin{equation}
 { U_D\over H_D}  = {E_{r+1(r+1)} \over H_{r+1}}
\label{maximal1}
\end{equation}
where the Lie algebra of the $U_D$--group $E_{r+1(r+1)} $ is the
maximally non compact real section of the exceptional $E_{r+1}$ series  of the simple complex
Lie Algebras
and $H_{r+1}$ is its maximally compact subalgebra \cite{cre}.
As   in the recent papers \cite{solvab1,solvab2},
the manifolds $E_{r+1(r+1)}/H_{r+1}$
share the distinctive  property of being non--compact homogeneous spaces of maximal rank
$r+1$, so that the associated solvable Lie algebras,
 such that ${E_{r+1(r+1)}}/{H_{r+1}} \, = \, \exp \left [ Solv_{(r+1)} \right ]
$,  have the particularly simple structure:
\begin{equation}
Solv\, \left ( E_{r+1}/H_{r+1} \right )\, = \, {\cal H}_{r+1} \, \oplus_{\alpha \in
\Phi^+(E_{r+1})} \, \IE^\alpha
\label{maxsolv1}
\end{equation}
where $\IE^\alpha \, \subset \, E_{r+1}$ is the 1--dimensional subalgebra associated
with the root $\alpha$
and $\Phi^+(E_{r+1})$ is the positive part of the $E_{r+1}$--root--system.
\par
The generators of the solvable Lie algebra  are in one to one
correspondence with the scalar fields of the theory.
Therefore they can be characterized as Neveu Schwarz or Ramond Ramond
depending on their origin in compactified string theory. From the
algebraic point of view the generators of the solvable algebra are of
three possible types:
\begin{enumerate}
\item {Cartan generators }
\item { Roots that belong to the adjoint representation of the
$D_r \equiv SO(r,r) \subset E_{r+1(r+1)}$ subalgebra (= the T--duality algebra) }
\item {Roots which are weights of an irreducible representation
 of the $D_r$ algebra.}
\end{enumerate}
The scalar fields associated with generators of type 1 and 2 in the above
list are Neveu--Schwarz fields while the fields of type 3 are
Ramond--Ramond fields.
\par
In the $r=6$ case, corresponding to $D=4$, there is one extra root,
besides those listed above, which is also of the Neveu--Schwarz type.
From the dimensional reduction viewpoint the origin of this extra
root is the following: it is associated with the axion $B_{\mu\nu}$
which only in 4--dimensions becomes equivalent to a scalar field.
This root (and its negative) together with the 7-th Cartan generator
of $O(1,1)$ promotes the S--duality in $D=4$ from $O(1,1)$, as it is in
all other dimensions, to $SL(2,\IR)$.

%%%%%%%%%%%%%%%%%%%%%%%%%%%%%%%%%%%%%%%%%%%%%%%%%%%%%%%%%%%%%%%%%%%%%

\subsection{Counting  of massless modes in sequential toroidal compactifications
of $D=10$ type IIA superstring}
In order to make the pairing between scalar field modes and solvable Lie algebra generators
explicit, it is convenient to organize the counting of bosonic zero modes
in a sequential way that goes down from $D=10$ to $D=4$ in 6 successive steps.

The useful feature of this sequential viewpoint is that it has a direct algebraic
counterpart in the successive embeddings of the exceptional Lie Algebras $E_{r+1}$
one into the next one:
{\footnotesize
\begin{equation}
  \matrix{E_{7(7)}&\supset &  E_{6(6)}&\supset & E_{5(5)}&\supset & E_{4(4)}&\supset
  & E_{3(3)}&\supset & E_{2(2)}&\supset & O(1,1) \cr
D=4 & \leftarrow & D=5 & \leftarrow & D=6 & \leftarrow & D=7 & \leftarrow & D=8 &
\leftarrow & D=9 & \leftarrow & D=10 \cr}
\end{equation}}
If we consider the bosonic massless spectrum \cite{gsw} of  type II theory in $D=10$
in the Neveu--Schwarz sector we have the metric, the axion and the dilaton,
while in the Ramond--Ramond sector we have a 1--form and a 3--form:
\begin{equation}
 D=10 \quad : \quad  \cases{
 NS: \quad g_{\mu\nu}, B_{\mu\nu} , \Phi \cr
 RR: \quad  A_{\mu} , A_{\mu\nu\rho} \cr}
 \label{d10spec}
\end{equation}
corresponding to the following counting of degrees of freedom:
$\# $ d.o.f. $g_{\mu\nu} = 35$, $\# $ d.o.f. $B_{\mu\nu} = 28$, $\# $ d.o.f. $A_{\mu} = 8$, $\# $ d.o.f. $A_{\mu\nu\rho} = 56$
so that the total number of degrees of freedom is $64$  both in the Neveu--Schwarz
and in the Ramond:
 \begin{eqnarray}
  \mbox{Total $\#$ of NS degrees of freedom}&=&{ 64}={ 35}+{ 28}+ { 1} \nonumber\\
  \mbox{Total $\#$ of RR degrees of freedom}&=&{ 64}={ 8}+{ 56}
  \label{64NSRR}
 \end{eqnarray}
 \par
It is worth noticing that the number of degrees of freedom of N--S and R--R sectors are equal, both for
bosons and fermions, to $128= (64)_{NS} + (64)_{RR}$. This is merely a consequence
of type II supersymmetry.
Indeed, the entire Ramond sector (both in type IIA and type IIB) can be thought as a spin $3/2$
multiplet of the second supersymmetry generator.
\par
Let us now organize the degrees of freedom as they appear after toroidal compactification
on a $r$--torus \cite{pope}:
\begin{equation}
{\cal M}_{10} = {\cal M}_{D-r} \, \otimes T_r
\end{equation}
Naming with Greek letters the world indices on the $D$--dimensional
space--time and with Latin letters the internal indices referring to
the torus dimensions we obtain the results displayed in Table \ref{tabu1} and number--wise we
obtain the counting of Table \ref{tabu2}:
\par
\vskip 0.3cm
\begin{table}[ht]
\begin{center}
\caption{Dimensional reduction of type IIA fields}
\label{tabu1}
 \begin{tabular}{|l|l|c|c|c|c|r|}\hline
\vline & \null & \vline & Neveu Schwarz & \vline & Ramond Ramond & \vline \\
 \hline
 \hline
\vline & Metric & \vline &  $g_{\mu\nu}$& \vline  & \null & \vline   \\ \hline
\vline & 3--forms & \vline &  \null & \vline  & $A_{\mu\nu \rho}$ & \vline \\ \hline
\vline & 2--forms & \vline   & $B_{\mu\nu}$ & \vline   & $A_{\mu\nu i}$ & \vline  \\ \hline
\vline & 1--forms & \vline   &  $g_{\mu i}, \quad B_{\mu i}$ & \vline  & $A_{\mu},
 \quad A_{\mu ij}$ & \vline \\ \hline
\vline & scalars  & \vline  & $\Phi, \quad g_{ij}, \quad B_{ij}$ &
 \vline  & $A_{i}, \quad A_{ijk}$ & \vline \\ \hline
 \end{tabular}
 \end{center}
\end{table}
 \par
 \vskip 0.3cm
 \par
\vskip 0.3cm
\begin{table}[ht]
\begin{center}
\caption{Counting of type IIA fields}

\label{tabu2}
 \begin{tabular}{|l|l|c|c|c|c|r|}\hline
\vline & \null & \vline & Neveu Schwarz & \vline & Ramond Ramond & \vline \\
 \hline
 \hline
\vline & Metric & \vline &  ${ 1}$& \vline  & \null & \vline   \\ \hline
\vline & $\#$ of 3--forms & \vline &  \null & \vline  & ${  1}$ & \vline \\ \hline
\vline &$\#$ of  2--forms & \vline   & ${  1}$ & \vline   & ${  r}$ & \vline  \\ \hline
\vline &$\#$ of 1--forms & \vline   &  $ {  2 r}$ & \vline  & $ {  1} +
\frac 1 2 \, r \, (r-1)$ & \vline \\ \hline
\vline & scalars  & \vline  & $1 \, +  \, \frac 1 2 \, r \, (r+1)$&
\vline  & $ r \, +\, \frac 1 6 \, r \, (r-1) \, (r-2)  $ & \vline \\
\vline &  \null   & \vline & $  +  \, \frac 1 2 \, r \, (r-1)  $ & \vline & \null   & \vline \\
\hline
\end{tabular}
\end{center}
\end{table}
\vskip 0.2cm
We can easily check that the total number
of degrees of freedom in both sectors is indeed $64$ after
dimensional reduction as it was before.

%%%%%%%%%%%%%%%%%%%%%%%%%%%%%%%%%%%%%%%%%%%%%%%%%%%
% SECTION  ALCHAIN  and DYNKIN %%%%%%%%%%%%%%%%%%%
%%%%%%%%%%%%%%%%%%%%%%%%%%%%%%%%%%%%%%%%%%%%%%%%
\section{ $E_{r+1}$ subalgebra chains and their string interpretation}
We can now inspect the algebraic properties of the solvable Lie algebras
$Solv_{r+1}$ defined by eq. \eqn{maxsolv1} and illustrate the match between
these properties and the physical properties of the sequential compactification.
\par
Due to the specific structure \eqn{maxsolv1} of a maximal rank solvable Lie algebra
every chain of {\it regular embeddings}:
\begin{equation}
E_{r+1} \, \supset \,K^{0}_{r+1} \, \supset \, K^{1}_{r+1}\, \supset \, \dots \, \supset \,
 K^{i}_{r+1}\, \supset \, \dots
\label{aletto1}
\end{equation}
where $K^{i}_{r+1}$ are subalgebras of the same rank and with
the same Cartan subalgebra ${\cal H}_{r+1}$ as
$E_{r+1}$ reflects into  a corresponding sequence of embeddings
of solvable Lie algebras and,
 henceforth, of  homogenous non--compact scalar manifolds:
\begin{equation}
E_{r+1}/H_{r+1} \, \supset \,K^{0}_{r+1}/Q^{0}_{r+1} \,\supset \,  \dots  \,\supset \,
K^{i}_{r+1}/Q^{i}_{r+1}
\label{caten1}
\end{equation}
which must be endowed with a physical interpretation.
In particular we can consider embedding chains such that \cite{witten}:
\begin{equation}
K^{i}_{r+1}= K^{i}_{r} \oplus X^{i}_{1}
\label{spacco}
\end{equation}
where $K^{i}_{r}$ is a regular subalgebra of $rank= r$ and $X^{i}_{1}$
is a regular subalgebra of rank one.
Because of the relation between the rank and the number
of compactified dimensions such chains clearly
correspond to the sequential dimensional reduction of either typeIIA (or B) or of M--theory.
Indeed the first of such regular embedding chains we can consider is:
\begin{equation}
K^{i}_{r+1}=E_{r+1-i}\, \oplus_{j=1}^{i} \, O(1,1)_j
\label{caten2}
\end{equation}
This chain simply tells us that the scalar manifold of
supergravity in dimension $D=10-r$ contains the
direct product of the supergravity scalar manifold  in dimension $D=10-r+1$
with the 1--dimensional moduli
space of a $1$--torus (i.e. the additional compactification radius one gets by making a further
step down in compactification).
\par
There are however additional embedding chains that originate from the different choices
of maximal
ordinary subalgebras admitted by the exceptional Lie algebra of the $E_{r+1}$ series.
\par
All the $E_{r+1}$ Lie algebras contain a subalgebra $D_{r}\oplus O(1,1)$ so
that we can write the chain \cite{solvab1,solvab2}:
\begin{equation}
K^{i}_{r+1}=D_{r-i}\, \oplus_{j=1}^{i+1} \, O(1,1)_j
\label{dueachain}
\end{equation}
As we discuss more extensively in the subsequent two sections, and we already anticipated,
 the embedding chain \eqn{dueachain}
corresponds to the decomposition of the scalar manifolds into submanifolds spanned by either
 N-S or  R-R fields, keeping moreover track of the way they originate at each level of the
sequential dimensional reduction. Indeed the N--S fields correspond to generators of the
solvable Lie algebra that behave as integer (bosonic) representations of the
\begin{equation}
D_{r-i} \, \equiv \, SO(r-i,r-i)
\label{subalD}
\end{equation}
while R--R fields correspond to generators of the solvable Lie algebra assigned to the spinorial
representation of the subalgebras \eqn{subalD}.
A third chain of subalgebras is the following one:
\begin{equation}
K^{i}_{r+1}=A_{r-1-i}\,\oplus  \, A_1 \, \oplus_{j=1}^{i+1} \, O(1,1)_j
\label{duebchain}
\end{equation}
and a fourth one is
\begin{equation}
K^{i}_{r+1}=A_{r-i}\,  \oplus_{j=1}^{i+1} \, O(1,1)_j
\label{elechain}
\end{equation}
The physical interpretation of the \eqn{duebchain}, illustrated in the next subsection, has its
origin in type IIB string theory. The same supergravity effective lagrangian can be viewed as
the result of compactifying either version of type II string theory. If we take the IIB
interpretation
the distinctive fact is that there is, already at the $10$--dimensional level a complex scalar
field $\Sigma$ spanning the non--compact coset manifold $SL(2,\IR)_U/O(2)$.
The $10$--dimensional U--duality
group  $SL(2,\IR)_U$ must therefore be present in all lower dimensions and it
corresponds to the addend
$A_1$ of the chain \eqn{duebchain}.
\par
The fourth chain \eqn{elechain} has its origin  in an M--theory interpretation or in a
 physical problem posed by the
$D=4$ theory.
\par
If we compactify the $D=11$ M--theory to $D=10-r$ dimensions using an $(r+1)$--torus $T_{r+1}$,
the flat metric on this is parametrized by the coset manifold $GL(r+1) / O(r+1)$.
The isometry group of the $(r+1)$--torus moduli space is therefore $GL(r+1)$ and its
Lie Algebra is $A_r + O(1,1)$, explaining the chain \eqn{elechain}.
Alternatively, we may consider the origin of the same chain from a $D=4$ viewpoint.
There
 the electric vector field strengths do not span an irreducible representation
of the U--duality group $E_7$ but sit together with their magnetic counterparts in the irreducible
fundamental ${\bf 56}$ representation.  An important question therefore is that of
establishing which subgroup $G_{el}\subset E_7$ has an electric action on the field strengths. The
answer is \cite{hull}:
\begin{equation}
G_{el} \, = \, SL(8, \IR )
\end{equation}
since it is precisely with respect to this subgroup that the fundamental ${\bf 56}$ representation
of $E_7$
splits into: ${\bf 56}= {\bf 28}\oplus {\bf 28}$. The Lie algebra of the electric subgroup is
$A_7 \, \subset \, E_7$ and it contains an obvious subalgebra $A_6 \oplus O(1,1)$.
The intersection
of this latter with the subalgebra chain \eqn{caten2} produces the electric chain \eqn{elechain}.
In other words, by means of equation \eqn{elechain} we can trace back in each upper dimension
which
symmetries will maintain an electric action also at the end point of the dimensional reduction
sequence,
namely also in $D=4$.
\par
We have  spelled out the embedding chains of subalgebras that are physically significant from
a string theory viewpoint. The natural question to pose now  is  how to understand their
algebraic
origin and how to encode them in an efficient description holding true sequentially in all
dimensions,
namely for all choices of the rank $r+1=7,6,5,4,3,2$. The answer is provided by reviewing the
explicit construction of the $E_{r+1}$ root spaces in terms of $r+1$--dimensional
euclidean vectors
\cite{gilmore}.
%%%%%%%%%%%%%%%%%%%%%%%%%%%%%%%%%%%%%%%%%%%%%%%%%%%%%%%%%%%%%%%
\subsection{Dynkin diagrams of the  $E_{r+1(r+1)}$ root spaces and structure
of the associated solvable algebras}
The root system  of type $E_{r+1(r+1)}$  can be described
for all values of $1\le r \le 6$ in the following way. As any other
root system it is a finite subset of vectors $\Phi_{r+1}\, \subset\, \IR^{r+1}$
such that $\forall \alpha ,\beta \, \in \Phi_{r+1}$ one has
$ \langle \alpha , \beta \rangle \, \equiv  2 (\alpha , \beta )/ (\alpha , \alpha) \,
\in \, \ZZ $ and such that $\Phi_{r+1}$ is invariant with respect to
the reflections generated by any of its elements. For an explicit
listing of the roots we refer the reader to \cite{solvab1,solvab2}. We just
recall that
the most efficient way to deal simultaneously with all the above root systems and
see the emergence of the above mentioned embedding chains is to embed them in the
largest, namely in the $E_7$ root space. Hence the various root systems $E_{r+1}$
will be represented by appropriate subsets of the full set of $E_7$ roots. In this
fashion for all choices of $r$ the $E_{r+1}$ are anyhow represented by 7--components
Euclidean vectors of length 2.
\par
Given a basis of seven simple roots $\alpha_1 , \dots \, \alpha_7$ whose
scalar products are those predicted by the $E_7$ Dynkin diagram:
\begin{eqnarray}
\alpha_1 =\left \{-\frac {1}{2},-\frac {1}{2},-\frac {1}{2}, -\frac {1}{2}, -\frac {1}{2},
-\frac {1}{2}, \frac{1}{\sqrt{2}}\right \}\nonumber\\
\alpha_2 = \left \{ 0,0,0,0,1,1,0 \right \}\nonumber\\
\alpha_3 = \left \{ 0,0,0,1,-1,0,0 \right \}\nonumber\\
\alpha_4 = \left \{ 0,0,0,0,1,-1,0 \right \} \nonumber\\
\alpha_5 = \left \{ 0,0,1,-1,0,0,0 \right \} \nonumber\\
\alpha_6 = \left \{ 0,1,-1,0,0,0,0 \right \} \nonumber\\
\alpha_7 = \left \{ 1,-1,0,0,0,0,0 \right \} \nonumber\\
\label{e7simple}
\end{eqnarray}
the embedding of chain \eqn{caten2} is easily described. By considering the subset of
$r$ simple roots
$\alpha_1 , \alpha_2 \, \dots \, \alpha_r$ we realize the Dynkin diagrams of type $E_{r+1}$.
Correspondingly,
the subset of all roots pertaining to the root system $\Phi(E_{r+1}) \, \subset \,
\Phi(E_7)$ can be explicitly found.
At each step of the sequential embedding one  generator of the $r+1$--dimensional
Cartan subalgebra
${\cal H}_{r+1}$ becomes orthogonal to the roots of the subsystem
$\Phi(E_{r})\subset\Phi(E_{r+1})$,
while the remaining $r$ span the Cartan subalgebra of $E_{r}$.
In order to visualize the other chains of subalgebras it is convenient to make two observations.
The first is to note that the simple roots selected in eq. \eqn{e7simple} are of two types: six
of them have integer components and span the Dynkin diagram of a $D_6 \equiv SO(6,6)$ subalgebra,
while the seventh simple root has half integer components and it is actually a spinor weight
with respect to this subalgebra. This observation leads to the embedding chain \eqn{dueachain}.
Indeed it suffices to discard one by one the last simple root to see the embedding of the
$D_{r-1}$ Lie algebra into $D_{r}\subset E_{r+1}$. As discussed in the next section $D_{r}$
is the Lie algebra of the T--duality group in type IIA toroidally compactified string theory.
\par
The next  observation is that the $E_7$ root system contains an exceptional pair of
roots $\beta =\pm \sqrt{2} \epsilon_7 \equiv \pm \sqrt{2} (0,0,0,0,0,0,1)$,
which does not belong to any of the other $\Phi (E_r)$
root systems. Physically the origin of this exceptional pair is very clear. It is associated
with the axion field $B_{\mu\nu}$ which in $D=4$ and only in $D=4$ can be dualized to an
additional scalar field. This root has not been chosen to be a simple root in eq.\eqn{e7simple}
since it can be regarded as a composite root in the $\alpha_i$ basis. However we have the
possibility
of discarding either $\alpha_2$ or $\alpha_1$ or  $\alpha_4$ in favour of $\beta$ obtaining a new
basis for the $7$-dimensional euclidean space $\IR^7$. The three choices in this operation
lead to the three different Dynkin diagrams given in fig.s (\ref{stdual}) and (\ref{elecal}),
corresponding to
the Lie Algebras:
\begin{equation}
 A_5 \oplus A_2\, , \quad   D_6\oplus A_1  \, , \quad
  A_7
\label{splatto}
\end{equation}

\iffigs
\begin{figure}
\caption{}
\label{stdual}
\epsfxsize = 10cm
\epsffile{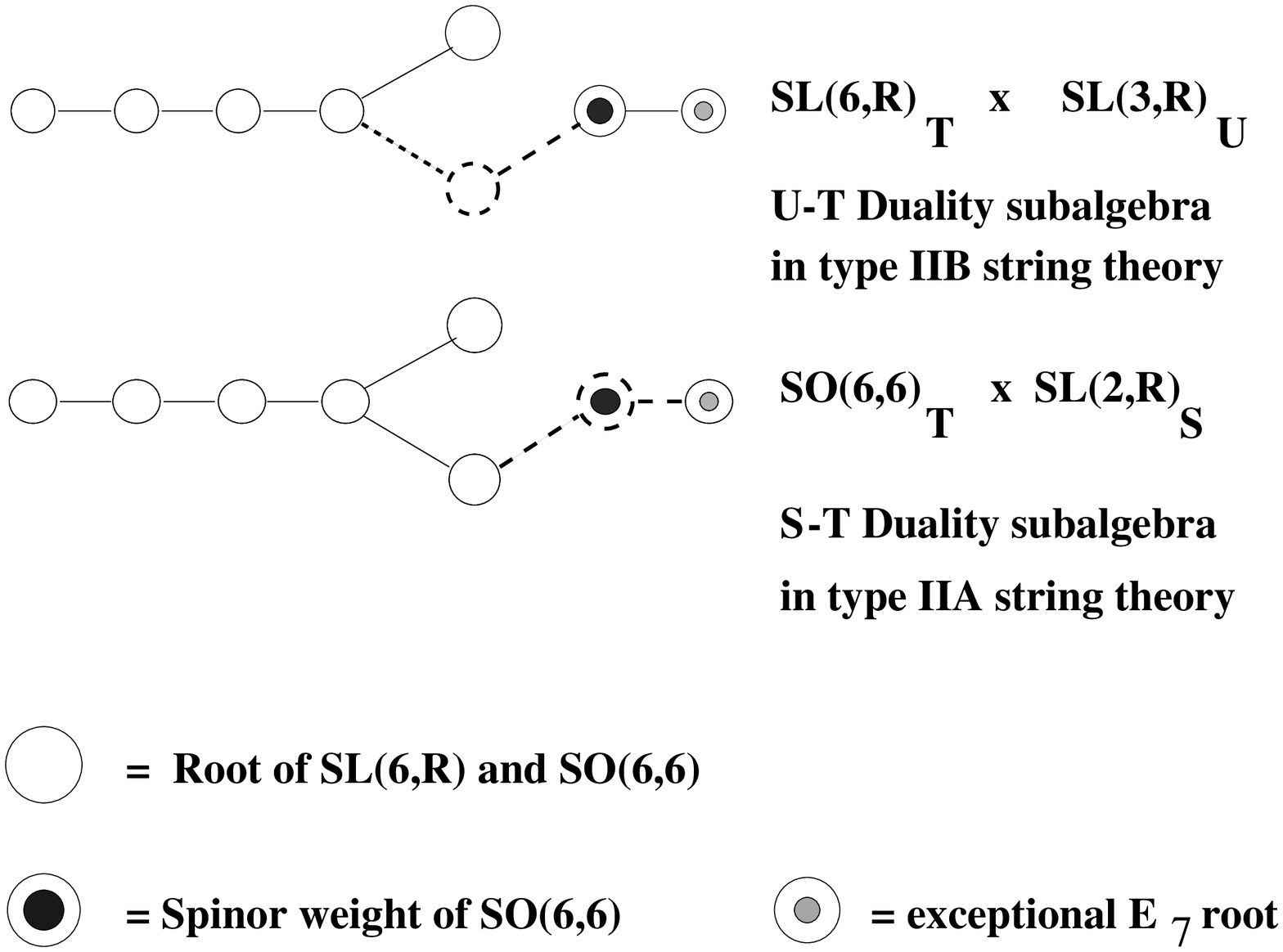}
\vskip -0.1cm
\unitlength=1mm
\end{figure}
\fi

\iffigs
\begin{figure}
\caption{}
\label{elecal}
\epsfxsize = 10cm
\epsffile{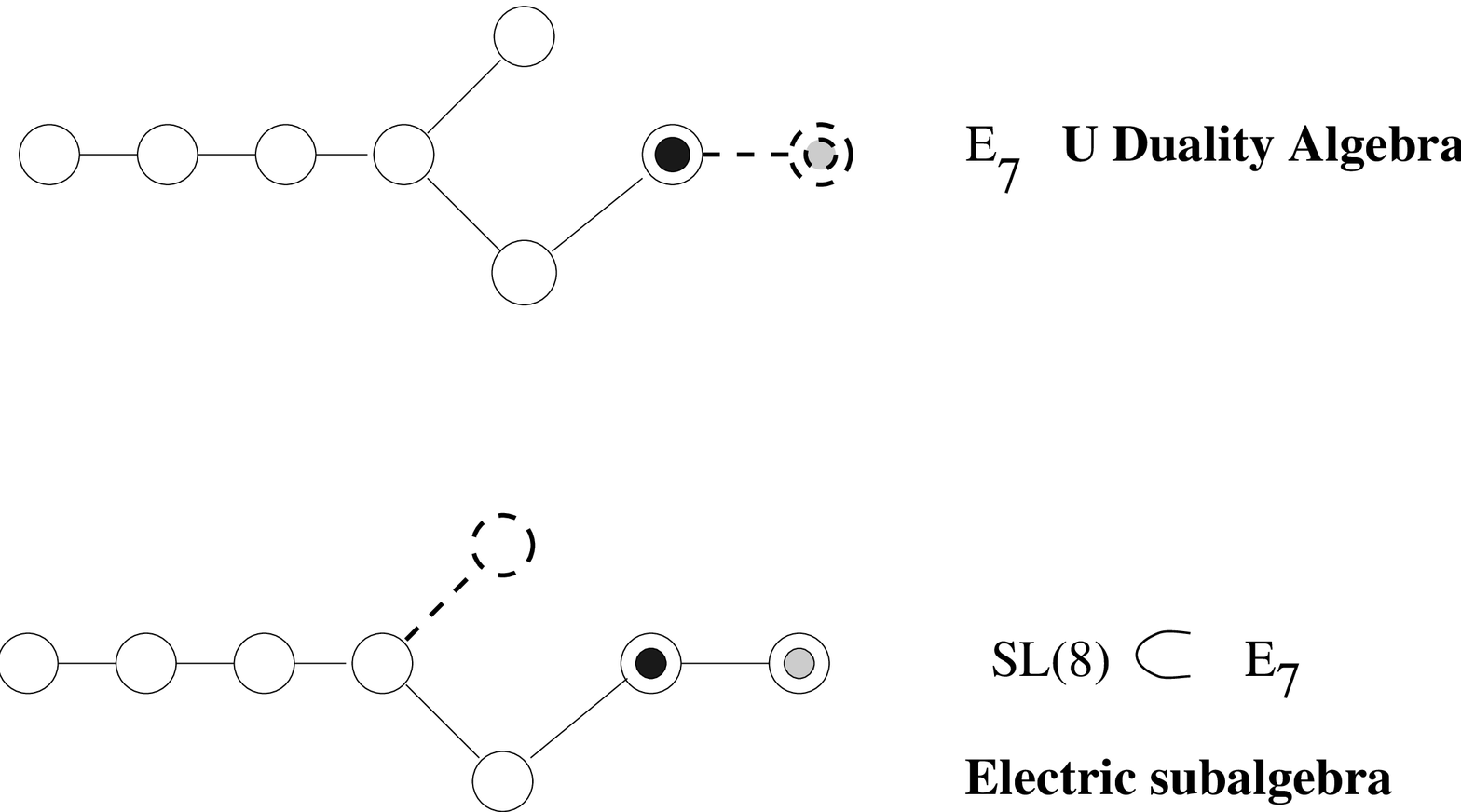}
\vskip -0.1cm
\unitlength=1mm
\end{figure}
\fi
From these embeddings occurring at the $E_7$ level, namely in $D=4$,
one deduces the three embedding chains
\eqn{dueachain},\eqn{duebchain},\eqn{elechain}: it just suffices to peal
off the last $\alpha_{r+1}$ roots
one by one and also the $\beta$ root that occurs only in $D=4$.
One observes that the appearance of the
$\beta$ root is always responsible for an enhancement of the S--duality group.
In the type IIA case
this group is enhanced from $O(1,1)$ to $SL(2,\IR)$ while in the type IIB case
it is enhanced from
the $SL(2,\IR)_U$ already existing in $10$--dimensions to $SL(3,\IR)$.
Physically this occurs by
combining the original dilaton field with the compactification radius of
the latest compactified
dimension.
%%%%%%%%%%%%%%%%%%%%%%%%%%%%%%%%%%%%%%%%%%%%%%%%%%%%%%%%%%%%
\subsection{String theory interpretation of the sequential embeddings:
Type $IIA$, type $IIB$ and $M$ theory chains}
We now turn to a closer analysis of the physical meaning of
the embedding chains we have been illustrating.
\par
Let us begin with the chain of eq.(\eqn{duebchain})that, as anticipated, is
related with the type IIB interpretation of supergravity theory.
The distinctive feature of this chain of embeddings is the presence
of an addend $A_1$ that is already present in 10 dimensions. Indeed
this $A_1$ is the Lie algebra of the $SL(2,R)_\Sigma $ symmetry of type $IIB$
D=10 superstring. We can name this group the U--duality symmetry $U_{10}$ in
$D=10$. We can use the chain \eqn{duebchain} to trace it in lower dimensions.
Thus let us  consider the decomposition
\begin{eqnarray}
E_{r+1(r+1)} & \rightarrow & N_r \otimes SL(2,\IR) \nonumber\\
N_r & = &   A_{r-1} \otimes O(1,1)
\la{sl2r}
\end{eqnarray}
Obviously $N_r$ is not contained in the  $T$-duality group $O(r,r)$ since the
$NS$ tensor field $B_{\mu \nu}$ (which  mixes with the metric under
$T$-duality) and the $RR$--field $B^c_{\mu \nu}$ form a doublet with
respect $SL(2,\IR)_U$. In fact, $SL(2,\IR)_U$ and  $O(r,r)$ generate
the whole U--duality group $E_{r+1(r+1)}$. The appropriate interpretation
of the normaliser of $SL(2,R)_\Sigma$ in $E_{r+1(r+1)}$  is
\begin{equation}
 N_r  = O(1,1) \otimes SL(r,\IR) \equiv GL(r,\IR)
 \label{agnosco}
\end{equation}
where $GL(r,\IR)$ is the isometry group of the  classical moduli
space for the $T_r$ torus:
\begin{equation}
 \frac{GL(r,\IR)}{O(r)}.
\end{equation}
The decomposition of the U--duality group appropriate for the type $IIB$ theory is
\be
E_{r+1} \rightarrow U_{10} \otimes GL(r,\IR) = SL(2,\IR)_U \otimes O(1,1) \otimes SL(r,\IR).
\la{sl2rii}
\ee
Note that since $GL(r,\IR) \supset O(1,1)^r$, this translates into $E_{r+1} \supset
SL(2,\IR)_U \otimes O(1,1)^r$.  (In Type $IIA$, the corresponding chain would
be $E_{r+1} \supset O(1,1) \otimes O(r,r) \supset  O(1,1)^{r+1}$.)  Note that while
$SL(2,\IR)$ mixes $RR$ and $NS$ states, $GL(r,\IR)$ does not. Hence we can write the following
decomposition for the solvable Lie algebra:
\bea
Solv \left( \frac{E_{r+1}}{H_{r+1}} \right) &=& Solv \left(\frac{GL(r,\IR)}{O(r)} \otimes
\frac{SL(2,\IR)}{O(2)} \right) + \left(\frac{\bf r(r-1)}{\bf 2}, {\bf 2} \right) \oplus {\bf X}
\oplus {\bf Y}  \nonumber \\
\mbox{dim }Solv \left( \frac{E_{r+1}}{H_{r+1}} \right)&=& \frac{d(3d-1)}{2} + 2 + x + y.
\la{solvii}
\eea
where $x=\mbox{dim }{\bf X} $ counts the scalars coming from the internal part of the $4$--form
$A^+_{{ \mu}{ \nu}{ \rho}{\sigma}}$ of type IIB string theory.
We have:
\begin{equation}
x =  \left \{
\matrix { 0 & r<4 \cr
\frac{r!}{4!(r-4)!} & r\geq 4 \cr}\right.
\la{xscal}
\end{equation}
and
\begin{equation}
y =\mbox{dim }{\bf Y} = \cases{
\matrix { 0 &  r  < 6   \cr 2 &  r = 6  \cr}\cr}.
\la{yscal}
\end{equation}
counts the scalars arising from dualising the two-index tensor
fields in $r=6$.
\par
For example, consider the $ D=6$ case. Here the type $IIB$  decomposition is:
\begin{equation}
E_{5(5)}=\frac{O(5,5)}{O(5) \otimes O(5)} \rightarrow \frac{GL(4,\IR)}{O(4)}
\otimes \frac{SL(2,\IR)}{O(2)}
\la{exii}
\end{equation}
whose compact counterpart is given by $O(10) \rightarrow SU(4) \otimes SU(2) \otimes U(1)$,
corresponding to the decomposition: ${\bf 45} =
{\bf (15,1,1)}+ {\bf (1,3,1)} + {\bf (1,1,1)} +{\bf (6,2,2)} + {\bf (1,1,2)}$. It follows:
\be
Solv(\frac{E_{5(5)}}{O(5) \otimes O(5)}) = Solv (\frac{GL(4,\IR)}{O(4)}
\otimes \frac{SL(2,\IR)}{O(2)}) +({\bf 6},{\bf 2})^+ + ({\bf 1},{\bf 1})^+.
\la{solvexii}
\ee
where the factors on the right hand side parametrize the internal part of the metric $g_{ij}$,
the dilaton and the $RR$ scalar ($\phi$, $\phi^c$), ($B_{ij}$, $B^c_{ij}$) and $A^+_{ijkl}$
respectively.
\par
There is a connection between the decomposition (\eqn{sl2r}) and the corresponding chains
in M--theory. The type IIB chain is given by eq.(\eqn{duebchain}),
namely by
\begin{equation}
E_{r+1(r+1)} \rightarrow SL(2,\IR) \otimes GL(r,\IR)
\end{equation}
 while the $M$ theory is given by eq.(\eqn{elechain}), namely by
 \begin{equation}
E_{r+1} \rightarrow O(1,1) \otimes SL(r+1,\IR)
\end{equation}
coming from the moduli space of $T^{11-D} = T^{r+1}$.
We see that these decompositions involve the classical moduli spaces of $T^r$
 and of $T^{r+1}$ respectively.
Type $IIB$ and $M$ theory decompositions
become identical if we decompose further $SL(r, \IR) \rightarrow O(1,1)
\times SL(r-1,\IR)$ on the type $IIB$ side and  $SL(r+1, \IR) \rightarrow O(1,1)
\otimes SL(2,\IR) \otimes SL(r-1,\IR)$ on the $M$-theory side. Then we obtain for both theories
\be
E_{r+1} \rightarrow SL(2,\IR) \times O(1,1) \otimes O(1,1) \otimes SL(r-1,\IR),
\la{sl2rall}
\ee
and we see that the group $SL(2,\IR)_U$ of type $IIB$ is identified with the
complex structure of the $2$-torus factor of the total
compactification torus $T^{11-D} \rightarrow T^2 \otimes T^{9-D}$.
\par
Note that according to  \eqn{splatto} in 8 and 4 dimensions, ($r=2$ and $6$)
in the decomposition \eqn{sl2rall} there is the following enhancement:
\begin{eqnarray}
&  SL(2,\IR) \times O(1,1)  \rightarrow SL(3,\IR) \quad (\mbox{for} \, r=2,6) & \\
 &\left\{\matrix{O(1,1) & \rightarrow & SL(2,\IR) \quad (\mbox{for} \, r=2) \cr
SL(5,\IR) \times O(1,1) & \rightarrow & SL(6,\IR) \quad (\mbox{for} \, r=6) \cr}\right. &
\end{eqnarray}
Finally, by looking at fig.(\ref{wite5}) let us observe that
$E_{7(7)}$ admits also a subgroup $SL(2,\IR)_T$ $\otimes
(SO(5,5)_S$ $\equiv E_{5(5)})$ where the $SL(2,\IR)$ factor is a
T--duality group, while the factor $(SO(5,5)_S$ $\equiv E_{5(5)})$
is an S--duality group which mixes RR and NS states.
\iffigs
\begin{figure}
\caption{}
\label{wite5}
\epsfxsize = 10cm
\epsffile{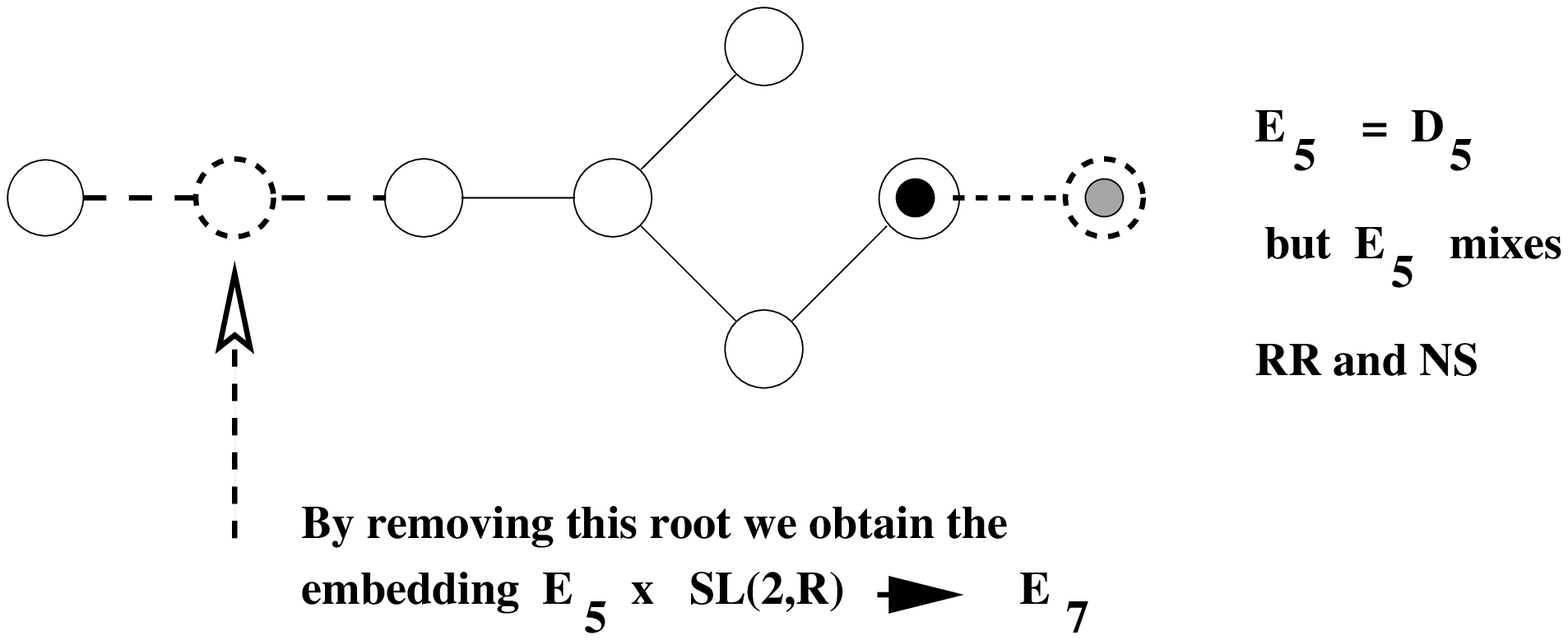}
\vskip -0.1cm
\unitlength=1mm
\end{figure}
\fi
%%%%%%%%%%%%%%%%%%%%%%%%%%%%%%%%%%%%%%%%%%%%%%%%%%%%%%%%%%%%%%%%%
\section{The maximal abelian ideals ${\cal A}_{r+1} \subset Solv_{r+1}$
of the solvable Lie algebra}

It is interesting to work out the maximal abelian ideals
${\cal A}_{r+1} \subset Solv_{r+1}$  of the
solvable Lie algebras generating the scalar manifolds
of maximal supergravity in dimension $D=10-r$.
The maximal abelian ideal of a solvable Lie
algebra is defined as the maximal subset of nilpotent generators  commuting among themselves.
From a physical point of view this is the largest abelian Lie algebra
that one might expect to be able to gauge in the supergravity theory. Indeed, as it turns
out, the number of vector fields in the theory is always larger
or equal than $\mbox{dim}{\cal A}_{r+1}$. Actually, as we are going to
see, the {\it gaugeable} maximal  abelian algebra
is always a proper subalgebra ${\cal A}^{gauge}_{r+1} \subset {\cal A}_{r+1}$
of this ideal.
\par
The criteria to determine ${\cal A}^{gauge}_{r+1}$ will be discussed
in the next section. In the present section we
derive ${\cal A}_{r+1}$ and we explore its relation with the
space of vector fields in one dimension above the dimension
we each time consider.  From such analysis we obtain a
filtration of the solvable Lie algebra which provides us
with a canonical polynomial parametrization of the supergravity
scalar coset manifold $U_{r+1}/H_{r+1}$
\par
%%%%%%%%%%%%%%%%%%%%%%%%%%%%%%%%%%%%%%%%%
\subsection{The maximal abelian ideal from an algebraic viewpoint}
Algebraically the maximal abelian ideal can be characterized by
looking at the decomposition of the U--duality algebra $E_{r+1(r+1)}$ with
respect to the U--duality algebra  in one dimension above.
In other words we have to consider the decomposition of $E_{r+1(r+1)}$ with
respect to the subalgebra $E_{r (r )} \, \otimes \, O(1,1)$. This
decomposition follows a general pattern which is given by the
next formula:
\begin{equation}
 \mbox{adj }  E_{r+1(r+1)} \, = \, \mbox{adj }  E_{r(r)}  \, \oplus
 \, \mbox{adj }  O(1,1) \, \oplus ( \ID^+_{r} \oplus \ID^-_{r} )
 \label{genpat}
\end{equation}
where $\ID^+_{r}$ is at the same time an irreducible representation of
the  U--duality algebra $E_{r (r )}$ in $D+1$ dimensions and coincides with the
maximal abelian ideal
\begin{equation}
\ID^+_{r}    \, \equiv \, {\cal A}_{r+1}   \, \subset \, Solv_{(r+1)}
\label{genmaxab}
\end{equation}
of the solvable  Lie algebra we are looking for. In eq. \eqn{genpat}
the subspace $\ID^-_{r}$ is just a second identical copy of the representation
 $\ID^{+}_{r}$ and it is made of negative rather than of positive weights
 of $E_{r (r )}$. Furthermore   $\ID^{+}_{r}$  and   $\ID^-_{r}$
 correspond to the eigenspaces belonging respectively to the eigenvalues
 $\pm 1$ with respect to the adjoint action of the S--duality group
 $O(1,1)$.
%%%%%%%%%%%%%%%%%%%%%%%%%%%%%%%%%%%%%%%%%%%%
\subsection{The maximal abelian ideal from a physical perspective: the vector fields in one
dimension above and translational symmetries}
Here, we would like to show that the dimension of the abelian
ideal in $D$ dimensions is equal to the number of vectors in dimensions $D+1$.
Denoting the number of compactified dimensions by $r$ (in string theory,
$r=10-D$), we will label the $U$-duality group in $D$ dimensions by $U_D
= E_{11-D} = E_{r+1}$. The $T$-duality group is $O(r,r)$, while the
$S$-duality group is $O(1,1)$ in  dimensions higher than four, $SL(2,R)$
in $D=4$ (and it is inside $O(8,8)$ in $D=3$).
\par
It follows from \eqn{genpat}
that the total dimension of the abelian ideal is given by
\be
{\rm dim} \, \cA_{D} \,\equiv \,  {\rm dim} \, \cA_{r+1}
 \,\equiv \,  {\rm dim} \,
\ID_{r}
\la{abeliii}
\ee
where $\ID_{r} $ is a representation of $U_{D+1}$ pertaining to the vector fields.
 According to \eqn{genpat} we have (for $D \ge 4$):
\be
\mbox{adj } U_D = \mbox{adj } U_{D+1} \oplus {\bf 1} \oplus ({\bf 2}, \ID_r).
\la{irrepu}
\ee
This is just an immediate consequence of the embedding chain \eqn{caten2}
which at the first level of iteration yields
$E_{r+1} \rightarrow E_r \times O(1,1)$. For example, under
$E_7 \rightarrow E_6 \times O(1,1)$ we have the branching rule:
${\rm adj} \, E_7 = {\rm adj} \, E_6 + {\bf 1} + ({\bf 2},{\bf 27})$ and
the abelian ideal is given by the ${\bf 27^+}$ representation of the $E_{6(6)}$ group.
The $70$ scalars of the $D=4, N=8$ theory are
naturally decomposed as ${\bf 70} = {\bf 42} +{\bf 1} +{\bf 27^+}$.
To see the splitting of
the abelian ideal scalars into $NS$  and $RR$ sectors, one has to consider
the decomposition of $U_{D+1}$ under the T--duality group $T_{D+1} = O(r-1, r-1)$,
namely the second iteration of the embedding chain \eqn{caten2}: $E_{r+1}
\rightarrow O(1,1) \times O(r-1,r-1)$. Then the vector  representation of $O(r-1, r-1)$
gives  the $NS$ sector, while the  spinor representation yields  the $RR$
sector. The  example of $E_7$ considered above is somewhat exceptional, since
we have ${\bf 27} \rightarrow ({\bf 10} + {\bf 1} +{\bf 16})$.
Here in addition to the expected ${\bf 10}$ and
${\bf 16}$ of $O(5,5)$ we find an extra $NS$ scalar:  physically this is due
to the fact that in
four dimensions the two-index antisymmetric tensor field  $B_{\mu \nu}$
is dual to a scalar, algebraically this  generator is associated with the
exceptional root $ \sqrt {2} \epsilon_7$.
To summarize, the $NS$ and $RR$ sectors are separately  invariant under
$O(r,r)$ in $D=10-r$ dimensions, while the abelian $NS$  and $RR$ sectors
are invariant under $O(r-1, r-1)$. The standard parametrization of
the
$U_D/H_D$ and $U_{D+1}/H_{D+1}$ cosets gives a clear illustration of this
fact:
\be
\frac{U_D}{H_D} \sim ( \frac{U_{D+1}}{H_{D+1}}, r_{D+1}, {\bf V}_r^{D+1}).
\la{cosetideal}
\ee
Here $r_{D+1}$ stands for the compactification radius, and ${\bf V}_r^{D+1} $
are the compactified vectors yielding the abelian ideal in $D$ dimensions.
\par
Note that:
\begin{equation}
  \mbox{adj}\, H_D = \mbox{adj} \, H_{D+1} \, +\,
\mbox{adj} \,\mbox{Irrep} \, U_{D+1}
\end{equation}
so it appears  that the abelian ideal forms a representation not only of
$U_{D+1}$ but also of the compact isotropy subgroup $H_{D+1}$ of the scalar coset
manifold.
\par
In the  above $r=6$ example we find
${\rm adj} \, SU(8) = {\rm adj} \, USp(8) \oplus {\bf 27^-}$,
$ \Longrightarrow $ ${\bf 63}= {\bf 36}+ {\bf 27^-}$.
%%%%%%%%%%%%%%%%%%%%%%%%%%%%%%%%%%%%%%%%%%%%%%%%%%%%%%%%
%%%%%%%%%%%%%%%%%%%%%%%%%%%%%%%%%%%%%%%%%%%%%%%%%%%%%%%%%%%%%%%%%%%%%%
%%%%%%%%%%%%%%%%%%%%%%%%%%%%%%%%%%%%%%%%%%%%%%%%%%%%%%%%

%%%%%%%%%%%%%%%%%%%%%%%%%%%%%%%%%%%%%%%%%%%%%%%%%%%%%%%%%%%%%%%%%%%%%%%%%%%%%%%%%
%%%%%%%%%%%%%%%%%%%%%%%%%%%%%%%%%%%%%%%%%%%%%%%%%%%%%%%%%%%%%%%%%%%%%%%%%%%%%%%%%%%%
%%%%%%%%%%%%%%%%%%%%%%%%%%%%%%%%%%%%%%%%%%%%%%%%%%%%%%%%%%%%%%%%%%%%%%%%%%%%%%%%%%%%
\section{Gauging}
In this last section we will consider the problem of gauging some isometries
of the coset $G/H$ in the framework of solvable Lie algebras.
\par
In particular we will consider in more detail the gauging of maximal compact groups
and the gauging of nilpotent abelian (translational) isometries.
\par
This procedure is a way of obtaining
partial supersymmetry breaking in extended supergravities \cite{hull},\cite{warn},\cite{huwa}
and it may find applications in the context of non perturbative phenomena in string
and M-theories.
\par
Let us consider the left--invariant 1--form $\Omega = L^{-1} dL$ of the coset
manifold $U_D/H_D$, where $L$ is the coset representative.
\par
The gauging procedure \cite{castdauriafre} amounts to the replacement of $dL$ with the gauge covariant
differential $\nabla L$ in the definition of the left--invariant 1--form $\Omega = L^{-1} d L$:
\begin{equation}
  \label{gaugedconn}
  \Omega  \rightarrow  \hat\Omega = L^{-1} \nabla  L =  L^{-1}( d + A )  L
= \Omega + L^{-1} A L
\end{equation}
As a consequence $\hat\Omega $ is no more a flat connection, but its curvature is given by:
\begin{equation}
  \label{gaugedchiusa}
 R(\hat\Omega)= d \hat\Omega +  \hat\Omega \wedge  \hat\Omega =  L^{-1}\cF  L \equiv  L^{-1}( dA + A\wedge A )  L
= L^{-1}(F^I T_I +L^I_{AB}T_I \bar\psi^A \psi^B )  L
\end{equation}
where $F^I$ is the gauged  supercovariant 2--form and $T_I$ are the generators of the gauge group
 embedded in the U--duality representation  of the vector fields.

 Indeed, by very definition, under the
full group $E_{r+1(r+1)}$ the gauge vectors are contained
in the representation $\ID_{r+1}$.
Yet, with respect to the gauge subgroup they must
transform in the adjoint representation, so that $\cG_D $ has to be chosen in such a way that:
\begin{equation}
\ID_{r+1} \, \stackrel{\cG_D}{\longrightarrow} \,\mbox{adj}\,\cG_D \, \oplus \, \mbox{rep}\,\cG_D
\label{brancione}
\end{equation}
where $\mbox{rep}\,\cG_D $ is some other representation of $\cG_D$ contained in the
above decomposition.

It is important to remark that vectors which are in $\mbox{rep} \cG_D$
(i.e. vectors which do not gauge $\cG_D$)
 may be required, by consistence
of the theory \cite{topiva}, to appear through their duals $(D-3)$--forms,
as for instance happens
for $D=5$ \cite{gurowa}.
In an analogous way $p$--form potentials ($p\neq 1$) which are in non trivial representations
of $\cG_D$ may also be required to appear through their duals $(D-p-2)$--potentials,
as is the case in $D=7$ for $p=2$ \cite{pepiva}.

The charges and the boosted structure constants discussed in the next
subsection can be retrieved from the two terms
appearing in the last expression of eq. \eqn{gaugedchiusa}

%%%%%%%%%%%%%%%%%%%%%%%%%%%%%%%%%%%%%%%%%%%%%%%%%%%%%%%%%%%%%%%%%%%%%%%%%%%%%%%%%%%%%%%
%%%%%%%%%%%%%%%%%%%%%%%%%%%%%%%%%%%%%%%%%%%%%%%%%%%%%%%%%%%%%%%%%%%%%%%%%%%%%%%%%%%%%%%
%%%%%%%%%%%%%%%%%%%%%%%%%%%%%%%%%%%%%%%%%%%%%%%%%%%%%%%%%%%%%%%%%%%%%%%%%%%%%%%%%%%%%%%
\subsection{Filtration of the $E_{r+1}$ root space, canonical
parametrization of the coset representatives and boosted structure constants}
As it has already been emphasized  in the introduction, the complete
structure of $N > 2$ supergravity in diverse dimensions is fully encoded in
the local differential geometry of the scalar coset manifold
$U_D/H_D$. All the couplings in the Lagrangian are described in terms
of the metric, the connection and the coset representative \eqn{cosrep1} of
$U_D/H_D$. A particularly significant consequence of extended
supersymmetry is that the fermion masses and the scalar potential the
theory can develop occur only as a consequence of the gauging and can
be extracted from a decomposition in terms of irreducible $H_D$ representations
of the {\sl boosted structure constants}\cite{mainaetal} \cite{castdauriafre}. Let us define these latter.
Let $\ID_{r+1}$ be the irreducible representation of the $U_D$
U--duality group pertaining to the vector fields and denote by ${\vec
{\bf w}}_\Lambda$ a basis for $\ID_{r+1}$:
\begin{equation}
\forall {\vec {\bf v}} \, \in \, \ID_{r+1}  \qquad : \qquad  {\vec v}\, = \,
v^\Lambda \, {\vec {\bf w}}_\Lambda
\label{baset}
\end{equation}
In the case we consider of maximal supergravity theories, where the
U--duality groups are given by  $E_{r+1(r+1)}$
 the basis vectors ${\vec {\bf w}}_\Lambda$ can be identified
with the $56$ weights of the fundamental $E_{7(7)}$ representation or with
the subsets of this latter corresponding to the irreducible
representations of its  $E_{r+1(r+1)}$ subgroups, according to the
branching rules:
\begin{equation}
{\bf 56}{\stackrel{E_6}{\longrightarrow}}\cases{{\bf 27 + 1}{\stackrel{E_5}{\longrightarrow}}
 \cases{{\bf 16}{\stackrel{E_4}{\longrightarrow}}{\bf \dots}\cr
 {\bf 10}{\stackrel{E_4}{\longrightarrow}}{\bf \dots}\cr
 {\bf 1+1}{\stackrel{E_4}{\longrightarrow}}{\bf \dots}\cr} \cr
 {\bf 27 + 1}{\stackrel{E_5}{\longrightarrow}}
 \cases{{\bf 16}{\stackrel{E_4}{\longrightarrow}}{\bf \dots}\cr
 {\bf 10}{\stackrel{E_4}{\longrightarrow}}{\bf \dots}\cr
 {\bf 1+1}{\stackrel{E_4}{\longrightarrow}}{\bf \dots}\cr} \cr
 }
\label{brancio}
\end{equation}
Let:
\begin{equation}
< ~, ~ > \quad : \quad \ID_{r+1} \, \times \ID_{r+1} \, \longrightarrow \,
\IR
\label{norma}
\end{equation}
denote the invariant scalar product in $\ID_{r+1}$ and let
$ {\vec {\bf w}}^\Sigma$ be a dual basis such that
\begin{equation}
 < {\vec {\bf w}}^\Sigma , {\vec {\bf w}}_\Lambda > \, = \,
 \delta^\Sigma_\Lambda
 \label{dualba}
\end{equation}
Consider then the $\ID_{r+1}$ representation of the coset representative
\eqn{cosrep1}:
\begin{equation}
L(\phi) \quad : \quad  \vert {\vec {\bf w}}_\Lambda >  \, \longrightarrow
L(\phi)_\Lambda^\Sigma \, \vert {\vec {\bf w}}_\Sigma >,
\label{matricia}
\end{equation}
and let $T^I$ be the generators of the gauge algebra $\cG_D \,
\subset E_{r+1(r+1)}$.

  The only admitted generators are those with
index $\Lambda = I \, \in \, {adj }\,\cG_D $,
and there are no gauge group generators with index $\Lambda \, \in \,\mbox{rep}\,\cG_D  $.
Given these definitions
the {\it boosted structure constants} are the following three--linear
$3$--tensors in the coset representatives:
\begin{equation}
\IC_{\Sigma\Gamma}^{\Lambda}\left(\phi\right)\, \equiv \,
\sum _{I=1}^{{\rm dim }\cG_D} \,
< {\vec {\bf w}}^\Lambda \, , \, L^{-1}\left(\phi\right) \, T_I \,
  L\left(\phi\right)  \, {\vec {\bf w}}_\Sigma > \, < {\vec {\bf w}}^I \,
 , \, L\left(\phi\right) {\vec {\bf w}}_\Gamma >
 \label{busto}
\end{equation}
and by decomposing them into irreducible $H_{r+1}$ representations we
obtain the building blocks utilized by supergravity in the fermion
shifts, in the fermion mass--matrices and in the scalar potential.
\par
In an analogous way, the charges appearing in the gauged
covariant derivatives are given by the following general form:
\begin{equation}
Q_{I\Sigma}^{\Lambda} \, \equiv \,
< {\vec {\bf w}}^\Lambda \, , \, L^{-1}\left(\phi\right) \, T_I \,
  L\left(\phi\right)  \, {\vec {\bf w}}_\Sigma >
 \label{buscar}
\end{equation}
\par
The coset representative $L\left(\phi\right)$ can be written in a canonical polynomial
parametrization which should give a
simplifying  tool in
mastering the scalar field dependence of all  physical relevant
quantities.
This includes, besides mass matrices, fermion shifts and scalar potential,
also the central charges
 \cite{amicimiei}.
\par
The alluded parametrization is precisely what the solvable Lie
algebra analysis produces.
\par
To this effect let us decompose
the solvable Lie algebra of $E_{7(7)}/SU(8)$ in a sequential way utilizing
eq. \eqn{genpat}. Indeed we can write the equation:
\begin{equation}
Solv(E_{7(7)})={\cal H}_7 \oplus \Phi^{+}(E_{7})
\label{decompo1}
\end{equation}
where $\Phi^{+}(E_{7})$ is the $63$ dimensional positive part of the $E_7$ root space.
By repeatedly using eq.  \eqn{genpat} we obtain:
\begin{equation}
\Phi^{+}(E_7)=\Phi^+(E_2) \oplus \ID^{+}_{2} \oplus \ID^{+}_{3}
\oplus \ID^{+}_{4} \oplus \ID^{+}_{5}  \oplus \ID^{+}_{6}
\label{decompo}
\end{equation}
where $ \Phi^+(E_2) $ is the one--dimensional root space of the
U--duality group in $D=9$ and $\ID^{+}_{r+1}$ are the weight-spaces
of the $E_{r+1}$ irreducible
representations to which the vector field in $D=10-r$
are assigned. Alternatively, as we have already explained, ${\cal A}_{r+2}
\equiv \ID^{+}_{r+1}$ are
the maximal abelian ideals of the U--duality group in $E_{r+2}$ in
$D=10-r-1$ dimensions.
\par
We can easily check that the dimensions sum appropriately as follows from:
 \begin{eqnarray}
\mbox{dim}\, \Phi^{+}(E_7)\, &=& { 63} \nonumber\\
\mbox{dim}\, \Phi^+(E_2)\, &=& { 1}\quad\quad~\mbox{dim}\,
\ID^{+}_{2}\, = { 3}\nonumber\\
\mbox{dim}\, \ID^{+}_{3}\, &=& { 6}\quad\quad~\mbox{dim}\,
\ID^{+}_{4}\, = { 10}\nonumber\\
\mbox{dim}\, \ID^{+}_{5}\, &=& { 16}\quad\quad\mbox{dim}\,
\ID^{+}_{4}\, = { 27}\nonumber\\
\label{ideadim}
\end{eqnarray}
Relying on eq. \eqn{decompo1},
\eqn{decompo} we can introduce a canonical set of scalar field
variables:
\begin{eqnarray}
\phi^i & \longrightarrow &  Y_i \,\in {\cal H} \quad \quad i=1,\dots \,
r \nonumber\\
\tau^i_{k} & \longrightarrow &  \,D^{(k)}_{i} \, \in  \ID_k \quad i=1,\dots \,
\mbox{dim }\,\ID_k  \quad (k=2,\dots,6) \nonumber\\
\tau_1   & \longrightarrow & \ID_1 \, \equiv  \, E_2
\label{canoni}
\end{eqnarray}
and adopting the short hand notation:
\begin{eqnarray}
 \phi \cdot  {\cal H} & \equiv & \phi^i  \,   Y_i  \nonumber\\
 \tau_k \cdot \ID_k & \equiv & \tau^i_{k}  \,D^{(k)}_{i} \nonumber\\
 \label{fucili}
\end{eqnarray}
we can write the coset representative for maximal supergravity in dimension
$D=10-r$ as:
\begin{eqnarray}
L & = & \exp \left [ \phi \cdot  {\cal H} \right ] \,  \prod_{k=1}^{r}
\, \exp \left[   \tau_k \cdot \ID_k \right ]
\end{eqnarray}
The relevant point is that, defining:
 \begin{equation}
S^i  \, \equiv \, \exp[\phi^i Y_i]
\end{equation}
all entries of the matrix $L$ are   polynomials   in the
$S^i, \tau^i_k, \tau_1$ ``canonical'' variables.
This follows from the fact that all the matrices $\tau_k \cdot \ID_k$ are
nilpotent (at most of order 4). Furthermore when the gauge group is chosen within the
maximal abelian ideal it is evident from the definition of the boosted
structure constants \eqn{busto} that they do not depend on the scalar
fields associated with the  generators of the same ideal. In such gauging one
has therefore {\it a flat direction} of the scalar potential for each
generator of the maximal abelian ideal.
\par
In the next section we turn to considering the possible gaugings more
closely.
%%%%%%%%%%%%%%%%%%%%%%%%%%%%%%%%
%%%%%%%%%%%%%%%%%%%%%%%%%%%%%%%%%%%%%%%%%%%%%%%
%%%%%%%%%%%%%%%%%%%%%%%%%%%%%%%%%%%%%%%%%%%%%%%%%%%%%%%%%%%%%%%%%%%%%%%%%%%%%%%%%
%%%%%%%%%%%%%%%%%%%%%%%%%%%%%%%%%%%%%%%%%%%%%%%%%%%%%%%%%%%%%%%%%%%%%%%%%%%%%%%
\subsection{Gauging of compact and translational isometries}
A necessary condition for the gauging of a subgroup $\cG_D\subset U_D$ is that the
representation of the vectors
$\ID_{r+1}$ must contain $\mbox{adj} \cG_D$.
Following this prescription, the list of maximal compact gaugings $\cG_D$ in any dimensions
 is obtained in  the third column of Table \ref{tabcomp}.
In the other columns we list the $U_D$-duality groups, their maximal compact subgroups
 and the left-over representations for vector fields.

\begin{table}[ht]
\caption{Maximal gauged compact groups}
\label{tabcomp}
\begin{center}
\begin{tabular}{|c|c|c|c|c|}
\hline
$D$ & $U_D$ & $H_D$ & $\cG_D$ & $\mbox{rep}\cG_D$ \\
\hline
\hline
9 & $SL(2,\IR) \times O(1,1)$ & $O(2)$ &  $O(2)$ & 2 \\
\hline
8 & $SL(3,\IR) \times SL(2,\IR)$ & $O(3) \times O(2)$   & $ O(3)$ & $3$  \\
\hline
7 & $SL(5,\IR)$ & $USp(4) $ &  $ O(5) \sim USp(4) $ & 0 \\
\hline
6 & $O(5,5)$ & $ USp(4) \times USp(4) $  &  $O(5)$ & $5+1$ \\
\hline
5 & $E_{6,(6)}$ & $USp(8)$ &  $ O(6) \sim SU(4) $ & $2\times 6$ \\
\hline
4 & $ E_{7(7)} $ & $SU(8)$ & $O(8)$ & 0 \\
\hline
\end{tabular}
\end{center}
\end{table}

\begin{table}[ht]
\caption{Transformation properties under $\cG_D$ of 2- and 3-forms}
 \label{tabrepba}
  \begin{center}
    \begin{tabular}{|c|c|c|}
\hline
$D$ &  $\mbox{rep} \, B_{\mu\nu}$  & $\mbox{rep} \, A_{\mu\nu\rho}$ \\
\hline
\hline
9 & 2 & 0 \\
\hline
8 & 3 & 0 \\
\hline
7 & 0 & 5 \\
\hline
6 & 5 & 0 \\
\hline
5 & $2\times 6$ & 0 \\
\hline
 \end{tabular}
  \end{center}
\end{table}

We notice that, for any $D$, there are $p$--forms ($p$ =1,2,3) which are charged under
the gauge group $\cG_D$.
Consistency of these theories requires that such forms become massive.
It is worthwhile to mention how this can occur in two variants of the Higgs mechanism.
Let us define the (generalized) Higgs mechanism for a $p$--form  mass generation through
the absorption
of a massless ($p-1$)--form (for $p=1$ this is the usual Higgs mechanism).
The first variant is the anti-Higgs mechanism for a $p$--form \cite{antihiggs},
which is its absorption by a massless ($p+1$)--form.
It is operating, for $p=1$, in $D=5,6,8,9$
for a sextet of $SU(4)$, a quintet of $SO(5)$, a triplet
of $SO(3)$ and a doublet of $SO(2)$, respectively.
The second variant is the self--Higgs mechanism \cite{topiva},
which only exists for $p = (D-1)/2$, $D= 4k-1$.
This is a massless $p$--form which acquires a mass through a
topological mass term and therefore it
becomes a massive ``chiral'' $p$--form.
The latter phenomena was shown to occur in $D=3$ and $7$.
It is amazing to notice that the representation assignments dictated by
$U$--duality for the various $p$--forms is
precisely that needed for consistency of the gauging procedure (see Table \ref{tabrepba}).
\par
The other compact gaugings listed in Table \ref{tabcomp} are the $D=4$
\cite{dwni} and $D=8$ cases \cite{sase2}.
\par
It is possible to extend the analysis of gauging semisimple groups
also to the case of solvable Lie groups \cite{fegipo}.
For the maximal abelian ideals of $Solv(U_D/H_D)$ this amounts
to gauge an $n$--dimensional subgroup of the translational
symmetries under which at least $n$ vectors are inert. Indeed the
vectors the set of vectors that can gauge an abelian algebra
(being in its adjoint representation) must be neutral under the
action of such an algebra.
We find that in  any dimension $D$ the dimension of this abelian
group $\mbox{dim} \cG_{abel}$ is given precisely by ${\rm dim} (\mbox{rep} \cG_D)$
which appear in the decomposition of $\ID_{r+1}$ under $O(r+1)$.
We must stress that this criterium gives a necessary but not sufficient
condition for the existence of
the gauging of an abelian isometry group,  consistent with supersymmetry.
\begin{table}[ht]
    \caption{Decomposition of fields in representations of the compact  group $\cG_D=O(11-D)$}
    \label{tabcompab}
    \begin{center}\begin{tabular}{|c|c|c|c|c|}
    \hline
    & vect. irrep & adj($O(11-D)$) & $\cA$  & $\mbox{dim} \cG_{abel}$  \\
    \hline
    $D = 9$ & $1+2$ & 1 & 1 & 1 \\
   \hline
     $D= 8$& $ 3+3$ & 3 & 3 & 3  \\ \hline
     $D=7$ & $6 + 4 $ & 6 & 6 & 4  \\
    \hline
     $D=6$ & $ 10+ 5 + 1 $ & 10 & 10 &  $5 + 1 $ \\
    \hline
     $D=5$ & $15+6+6$ & 15 & $ 15+1$ &   $6+6$  \\
     \hline
     $D=4$ & $(21+7)\times 2$ & 21 & $21 + 1 \times 6$ & 7 \\
     \hline
    \end{tabular}\end{center}\end{table}
 \vskip 0.2cm
As already remarked in the introduction it is reasonable to expect
that the gauging of translational isometries and its consequence,
namely the spontaneous breaking of supersymmetry, should be generated
in the effective quantum lagrangian by the condensation of field
strengths. This will occur through the summation on non perturbative
states, namely on BPS monopoles, black-holes and $p$-branes.
\par
Hence, in the next section, I turn to discuss these topics.
%%%%%%%%%%%%%%%%%%%%%%%%%%%%%%%%%%%%%%%%%%%%%%%%
\section{BPS States in rigid N=2 supersymmetry}
\label{BPSrigida}
The most general form of a rigid N=2 super Yang--Mills Lagrangian was derived
in \cite{jgpnoi}: its structure is fully determined by three geometrical data:
\begin{itemize}
\item {The choice of a Special K\"ahler manifold of the rigid type
$ {\cal SK}^{rig}$ describing the vector multiplet couplings}
\item {The choice of a HyperK\"ahler manifold $ {\cal HK}$ describing
the hypermultiplet dynamics }
\item{The choice of a gauge group
$G^{gauge}\subset G^{iso}$,
subgroup of the isometry group of both $ {\cal SK}^{rig}$ and  $ {\cal HK}$ }
\end{itemize}
The bosonic action has the following form:
\begin{eqnarray}
 {\cal L}_{N=2\, SUSY}^{Bose} & =   &  g_{i {{j}^\star}}\,
\nabla^{\mu} z^i \nabla _{\mu} \bar z^{{j}^\star}\,+\,
h_{uv} \, \nabla _{\mu}\, q^u \, \nabla^{\mu}\, q^v -  \, {\rm V}\bigl ( z, {\bar
z}, q \bigr )
\nonumber\\
& & + \,{\rm i} \,\left(
\bar {\cal N}_{IJ} {\cal F}^{- I}_{\mu \nu}
{\cal F}^{- J\vert {\mu \nu}}
\, - \,
{\cal N}_{IJ} {\cal F}^{+ I}_{\mu \nu}
{\cal F}^{+ J \vert {\mu \nu}} \right )
\label{susaction}
\end{eqnarray}
where the scalar potential is expressed in terms of the killing
vectors $k^i_{I}$, $k^u_{I}$ generating the gauge group algebra on the scalar manifold
${\cal SK}\otimes {\cal HQ}$, of the upper
half $Y^{J}(z)$ of the symplectic section of rigid special geometry and also in terms
of the {\it momentum map} functions ${\cal P}^x_I(q)$ yielding the Poissonian realization
of the gauge group algebra on the HyperK\"ahler manifold:
\begin{eqnarray}
   \,-{\rm V}\bigl ( z, {\bar
z}, q \bigr )& = &
    -\,g^2\,\Bigl [ \left( g_{i{{j}^\star}} \, k^i_{I}\,k^{{j}^\star}_{J} +\,4\,h_{uv}
k^u_{I}\,k^v_{J} \right)\,
   \bar Y^{I}\,Y^{J}
   +\,g^{i{{j}^\star}}\,f^{I}_i\,f^{J}_{{j}^\star}\,
{\cal P}^x_{I}\,{\cal P}^x_{J}\,
 \Bigr] \nonumber\\
\label{susypot}
\end{eqnarray}
The kinetic term of the vectors in \eqn{susaction} involves the
period matrix  ${\cal N}_{IJ}$, which is also a datum
of rigid special geometry.
\par
If we restrict our attention to a pure gauge theory without
hypermultiplets, and we calculate the energy of a generic {\it static configuration}
({\it i.e} $F^I_{0a}$ = $0$,  $\nabla_0 z^i$ = $0$), we obtain:
\begin{eqnarray}
  E &=& \int \, d^3x \Bigl [  \mbox{Im}{\cal N}_{IJ} F^I_{ab}\,  F^J_{ab}
 + \, g_{i{{j}^\star}} \, \nabla_a z^i \,
\nabla_a {\bar z}^{j^\star} \, + \,g^2\, g_{i{{j}^\star}} \, k^i_{I}\,k^{{j}^\star}_{J}\,
   \bar Y^{I}\,Y^{J} \Bigr ] \nonumber\\
\label{monenergy}
\end{eqnarray}
Using the special geometry identity:
\begin{equation}
U^{IJ} \equiv g^{i{{j}^\star}} \, {\bar f}^J_{j^\star} \, f^I_i \, =
\,-\, \frac{1}{2} \, \left ( \mbox{Im} {\cal N} \right)^{-1\vert IJ}
\label{rigident}
\end{equation}
the energy integral \eqn{monenergy} can be rewritten according to a
{\it Bogomolny decomposition} as follows:
\begin{eqnarray}
E &=& \int \, d^3x  \, \frac{1}{2}\, g_{i{{j}^\star}}
\left( 2\mbox{i}\, G^i_{ab} \pm \epsilon_{abc} \nabla_c z^i \right)
\, \left(- 2\mbox{i}\, G^{j^\star}_{ab} \pm \epsilon_{abc} \nabla_c z^{j^\star} \right)
\label{1stcondi} \\
& & + \int \, d^3x  \, \,g^2\, g_{i{{j}^\star}} \, k^i_{I}\,k^{{j}^\star}_{J}\,
   \bar Y^{I}\,Y^{J}   \label{2ndcondi}  \\
 & &\pm \int \, d^3x  \, \mbox{i} \epsilon_{abc} \left(G^{j^\star}_{ab} \nabla_c z^{i}
 \, - \,  G^{i}_{ab} \nabla_c z^{j^\star} \right)
 \label{topocharge}
\end{eqnarray}
where, by definition:
\begin{equation}
G^{i}_{\mu\nu} \, \equiv \, g^{i{{j}^\star}}
\,{\bar f}^I_{j^\star} \,\mbox{Im}{\cal N}_{IJ} F^J_{\mu\nu}   \quad
; \quad
{\bar f}^I_{j^\star} \equiv \nabla_{j^\star}{\bar Y}^I
\end{equation}
The last \eqn{topocharge} of the three addends contributing to the energy is
the integral of a total divergence and can be identified with the
topological charge of the configuration:
\begin{eqnarray}
& Z \, \equiv \, 2 \int_{S_\infty^2} \,\mbox{Im}{\cal N}_{IJ} F^I \, \mbox{Im} Y^J & \nonumber\\
& = 2 \int_{R^3} \, \mbox{Im}{\cal N}_{IJ} F^I \, \wedge \nabla \mbox{Im} Y^J & \nonumber \\
& = \int \, d^3x  \, \mbox{i} \epsilon_{abc} \left(G^{j^\star}_{ab} \nabla_c z^{i}
 \, - \,  G^{i}_{ab} \nabla_c z^{j^\star} \right)\, g_{ij^\star}&
 \label{centratopo}
\end{eqnarray}
where $S_\infty^2$ is the $2$--sphere at infinity bounding  a constant time
slice of space--time. Since the other two addends \eqn{1stcondi},\eqn{2ndcondi} to the energy
of the static configuration are integrals of perfect squares, it follows that in each
topological sector, namely at fixed value of the topological charge
$Z$ the mass satisfies the Bogomolny bound \eqn{bogobound}.
Furthermore a BPS saturated  state (monopole or dyon) is defined by
the two conditions:
\begin{eqnarray}
 &2\mbox{i}\, G^i_{ab} \pm \epsilon_{abc} \nabla_c z^i = 0& \label{BPScondi1}\\
 & g^2\, g_{i{{j}^\star}} \, k^i_{I}\,k^{{j}^\star}_{J}\,
   \bar Y^{I}\,Y^{J} = 0 &
\label{BPScondi2}
\end{eqnarray}
The relation with the preservation of $\frac{1}{2}$ supersymmetries can now be easily seen.
In a bosonic background the supersymmetry variation of the bosons is
automatically zero since it is proportional to the fermion fields
which are zero: one has just to check the supersymmetry variation of
the fermion fields. In the theory under consideration the only
fermionic field is the gaugino and its SUSY variation is given by
(see \cite{jgpnoi}):
\begin{eqnarray}
  \delta\lambda^{iA}   &=  &  \mbox{i} \nabla_\mu z^i \, \gamma^\mu \, \epsilon^A
  + \varepsilon^{AB} \left( G^{-i}_{\mu\nu} \, \gamma^{\mu\nu} + k^i_I {\bar Y}^I \right) \,
\epsilon_B
\label{gaugvari}
\end{eqnarray}
If we use a SUSY parameter subject to the condition:
\begin{equation}
\gamma_0 \, \epsilon^A \, = \, \pm \mbox{i} \, \varepsilon ^{AB} \,
\epsilon_B
\label{parcondicio}
\end{equation}
then, in a static bosonic background eq.\eqn{gaugvari} becomes:
\begin{eqnarray}
  \delta\lambda^{iA} & =  &
  \Bigl [ -\mbox{i} \,\frac{1}{2} \, \left( 2 \mbox{i}\, G^i_{ab} \pm
\epsilon_{abc} \, \nabla_c z^i \right) \, \gamma^{ab} \, \epsilon_B \,
+ k ^i_I {\bar Y}^I \, \epsilon_B \, \Bigr ]\, \varepsilon^{AB} \,
\label{varispec}
\end{eqnarray}
Henceforth the configuration is invariant under the supersymmetries
of type \eqn{parcondicio} if and only if eq.\eqn{BPScondi1} is
satisfied together with:
\begin{equation}
 k^i_I {\bar Y}^I \, = \, 0
 \label{sqrtcondi}
\end{equation}
Eq.\eqn{sqrtcondi} is nothing else but the square--root of eq.\eqn{BPScondi2}.
So we can conclude that the BPS saturated states are just those
configurations which are invariant under supersymmetries of type
\eqn{parcondicio}. On the other hand, these supersymmetries are,
by definition, those generated by
the operators \eqn{redchar}. So by essential use of the {\it
rigid special geometry} structure we have shown the match between
the abstract reasoning of section ~\ref{centcharge} and the
concrete field theory realization of BPS saturated states.
\section{BPS black holes in  N=2 local supersymmetry}
\label{BPSlocala}
Eq.\eqn{parcondicio} is not Lorentz invariant and introduces a
clear--cut separation between space and time. The interpretation of
this fact is that we are dealing with localized lumps of energy that can be interpreted as
quasi--particles at rest. A Lorentz boost simply puts such quasi--particles
into motion. In the gravitational case the generalization of eq.\eqn{parcondicio}
requires the existence of a time--like killing vector $ \xi^\mu$, in order
to write:
\begin{equation}
 \xi^\mu \,\gamma_\mu \, \epsilon^A \, = \, \pm \mbox{i} \, \varepsilon ^{AB} \,
\epsilon_B
\label{lparcondicio}
\end{equation}
Furthermore, the analogue of the localization condition corresponds
to the {\it asymptotic flatness} of space--time.
\par
We are therefore led to look for the BPS saturated states of local $N=2$
supersymmetry within the class of {\it electrically and magnetically charged,
asymptotically flat, static space--times}. Generically such
space--times are black--holes since they have singularities hidden by horizons.
  Without the constraints imposed by supersymmetry the
horizons can also disappear and there exist configurations that display
{\it naked singularities}. In the supersymmetric case, however, the
Bogomolny bound \eqn{bogobound} becomes the statement that the ADM
mass of the black--hole is always larger or equal than the central charge.
This condition just ensures that the horizon exists. Hence the
{\it cosmic censorship} conjecture is just a consequence of $N\ge 2$
supersymmetry. This was noted for the first time in \cite{kalvanp1}.
The BPS saturated black--holes are configurations for
which the horizon area is minimal at fixed electric and magnetic
charges. This result was obtained by Ferrara and Kallosh in
\cite{ferkal4,ferkal2}.They are determined by solving the gravitational analogue
of the Bogomolny first order equations \eqn{BPScondi1}, \eqn{sqrtcondi},
obtained from the SUSY variation of the fermions.
\par
If we restrict our attention to the gravitational coupling of vector
multiplets, the bosonic action we have to consider is the
following one:
 \begin{eqnarray}
 S^{Bose}_{N=2} &=&
 \int \sqrt{-g}\,d^4 \,x  {\cal L}  \nonumber \\
  {\cal L}    &=&
  - {1 \over 2} \, R \,  + \, g_{i {{j}^\star}}\,
\nabla^{\mu} z^i \nabla _{\mu} \bar z^{{j}^\star} +  \,{\rm i} \,\left(
\bar {\cal N}_{\Lambda \Sigma} {\cal F}^{- \Lambda}_{\mu \nu}
{\cal F}^{- \Sigma \vert {\mu \nu}}
\, - \,
{\cal N}_{\Lambda \Sigma} {\cal F}^{+ \Lambda}_{\mu \nu}
{\cal F}^{+ \Sigma \vert {\mu \nu}} \right )
 \nonumber    \\
& & -\,g^2\, g_{i{{j}^\star}} \, k^i_{\Lambda}\,k^{{j}^\star}_{\Sigma}
\, \bar L^{\Lambda}\,L^{\Sigma}
\label{sugraction}
\end{eqnarray}
where $L^{\Lambda}$ denotes the upper half of a covariantly
holomorphic section of {\it local special geometry} and
${\cal N}_{\Lambda \Sigma}$ is the period matrix according to
its local rather than rigid definition (see \cite{mylec},\cite{jgpnoi}).
According to the previous discussion we consider for the metric an ansatz
of the following form:
\begin{equation}
ds^2 \, = \, e^{2U(r)} \, dt^2 \, - \, e^{-2U(r)} d{\vec x}^2
\label{umetric}
\end{equation}
where ${\vec x}$ are isotropic coordinates on $\IR^3$ and
$U(r)$ is a function only of:
\begin{equation}
r \equiv \sqrt{{\vec x}^2}
\label{rfun}
\end{equation}
As we shall see in the next section, eq.\eqn{umetric} corresponds to
a 0--brane ansatz. This is in line with the fact that we have 1--form
gauge fields in our theory that couple to 0--branes, namely to particle
world--lines. Indeed, in order to proceed further we need an ansatz for
the gauge field strengths. To this effect we begin by constructing a
2--form which is {\it anti--self--dual} in the background of the
metric \eqn{umetric} and whose integral on the $2$--sphere at
infinity $ S^2_\infty$ is normalized to $ 2 \pi $. A short
calculation yields:
\begin{eqnarray}
E^-   &=&   \mbox{i} \frac{e^{2U(r)}}{r^3} \, dt \wedge {\vec x}\cdot
d{\vec x}  +  \frac{1}{2}   \frac{x^a}{r^3} \, dx^b   \wedge
dx^c   \epsilon_{abc} \nonumber\\
2 \, \pi  &=&    \int_{S^2_\infty} \, E^-
\label{eaself}
\end{eqnarray}
and with a little additional effort one obtains:
\begin{equation}
E^-_{\mu\nu} \, \gamma^{\mu\nu}  = 2 \,\mbox{i} \frac{e^{2U(r)}}{r^3}  \,
\gamma_a x^a \, \gamma_0 \, \frac{1}{2}\left[ {\bf 1}+\gamma_5 \right]
\label{econtr}
\end{equation}
which will prove of great help in the unfolding of the supersymmetry
transformation rules. Then utilizing eq.\eqn{eaself} we write the
following ansatz for the gauge field--strengths:
\begin{eqnarray}
F^{\Lambda -}_{\mu\nu} & \equiv & \frac{1}{2} \left(F^{\Lambda }_{\mu\nu} -
\mbox{i}\,\frac{1}{2}\, \epsilon_{\mu\nu\rho\sigma} \,F^{\Lambda \vert\rho\sigma }
\right)  =  \frac{1}{4 \pi}\, t^\Lambda \, E^-_{\mu\nu} \nonumber\\
 t^\Lambda & = & \mbox{complex number}  =
 \int_{S^2_\infty}\,  F^{\Lambda -}_{\mu\nu} \, dx^\mu \wedge \, dx^\nu
 \label{gaugansz}
\end{eqnarray}
 Following the standard definitions occurring in the discussion of electric--magnetic
 duality rotations \cite{mylec} (and \cite{jgpnoi}) we also obtain:
\begin{eqnarray}
G^{-}_{\Lambda\vert \mu\nu}& = &{\bar {\cal N}}_{\Lambda\Sigma}
\, F^{\Sigma -}_{\mu\nu}  =
\frac{1}{4 \pi} \,{\bar {\cal N}}_{\Lambda\Sigma}\, t^\Sigma \, E^-_{\mu\nu}
\label{Gform}
\end{eqnarray}
To our purposes the most important field strength combinations are
the gravi--photon and matter--photon combinations occurring, respectively
in the gravitino and gaugino SUSY rules. They are defined by
(see \cite{jgpnoi}):
\begin{eqnarray}
 T^{-}_{\mu\nu}& =& 2\mbox{i}\, \left( \mbox{Im}
 {\cal N}\right)_{\Lambda\Sigma} \, L^\Lambda \,F^{\Sigma -}_{\mu\nu}
 \label{gravifot} \\
 G^{\pm i}_{\mu\nu} & = & -g^{ij^\star} \, {\bar f}^\Gamma_{j^\star}
 \, \left( \mbox{Im} {\cal N}\right)_{\Gamma\Lambda} \,F^{\Sigma \pm}_{\mu\nu}
 \label{matfot}
\end{eqnarray}
The central charge is defined by the integral of the graviphoton
\eqn{gravifot} (see \cite{cereferpro}):
\begin{equation}
Z \equiv \int_{S^2_\infty} \,T^{-}_{\mu\nu} \, dx^\mu \wedge dx^\nu
\label{zdef}
\end{equation}
Using eq.\eqn{gaugansz} and \eqn{gravifot}we obtain:
\begin{equation}
Z= 2\mbox{i}\, \left( \mbox{Im}
 {\cal N}\right)_{\Lambda\Sigma} \, L^\Lambda \, t^\Sigma
 \label{espres}
\end{equation}
while utilizing the identities of special geometry we also obtain:
\begin{equation}
T^{-}_{\mu\nu} = M_\Sigma \, F^\Sigma_{\mu\nu} - L^\Lambda \,
G_{\Lambda \vert \mu\nu}
\label{passaggio}
\end{equation}
where $M_\Sigma(z)$ is the lower part of the symplectic section of local
special geometry. Consequently we obtain:
\begin{eqnarray}
Z= M_\Sigma \, p^\Sigma - L^\Lambda \, q_\Lambda
\label{holcharge}
\end{eqnarray}
having defined the moduli dependent electric and magnetic
charges as follows:
\begin{eqnarray}
q_\Lambda & \equiv & \int_{S^2_\infty} \,G_{\Lambda\vert \mu\nu}\,
dx^\mu \wedge dx^\nu
\label{holele}\\
p^\Sigma & \equiv & \int_{S^2_\infty} \,F^{\Sigma}_{ \mu\nu}\,\,
dx^\mu \wedge dx^\nu
\label{holmag}
\end{eqnarray}
Alternatively, if following J. Schwarz \cite{sumschwarz} we define the
electric and magnetic charges by the asymptotic behaviour  of the
bare electric and magnetic fields:
\begin{equation}
F^{\Lambda}_{0a} \cong \frac{q^\Lambda_{(el)}}{r^3}\, x^a \quad ; \quad
{\tilde F}^{\Lambda}_{0a} \cong \frac{q^\Lambda_{(mag)}}{r^3}\, x^a
\label{elemag}
\end{equation}
we find the relations
\begin{equation}
q^\Lambda_{(el)}= 2\, \mbox{Im}\, t^\Lambda \quad ; \quad  q^\Lambda_{(mag)}
= 2\, \mbox{Re}\, t^\Lambda
\label{relazia}
\end{equation}
and
\begin{eqnarray}
t^\Lambda&=&\frac{1}{2}\,\Biggl \{ p^\Lambda+\mbox{i}\left(\mbox{Im} {\cal
N}_\infty^{-1}\right)^{\Lambda\Sigma}  \, \left[\left(\mbox{Re}{\cal
N}\right)_{\Sigma\Gamma}\, p^\Gamma - q_\Sigma\right]\Biggr \}
\label{relaziona}
\end{eqnarray}
In a fully general bosonic background the $N=2$ supersymmetry
transformation rules of the gravitino and of the gaugino are:
\begin{eqnarray}
\delta\,\psi_{A \vert \mu} &=& \nabla_\mu \epsilon_A -
\frac{1}{4} T^-_{\rho\sigma} \gamma^{\rho\sigma}
\, \gamma _\mu \,\epsilon^B \, \varepsilon ^{AB} \label{gravirule}\\
\delta\lambda^{\alpha A} & = & \mbox{i} \, \nabla_\mu z^\alpha \,
\gamma^\mu \, \epsilon^A + G^{-\alpha}_{\rho\sigma}
\gamma^{\rho\sigma} \epsilon^B \, \varepsilon ^{AB}
\nonumber\\
&&+ \varepsilon ^{AB} \, k^\alpha_\Lambda \, {\bar L}^\Lambda \,
\epsilon_B \label{gaugirule}
\end{eqnarray}
where the derivative:
\begin{equation}
\nabla_\mu \epsilon_A   \equiv   \Bigl (\partial_\mu -\frac{1}{4}
\omega^{ab}_\mu \, \gamma_{ab}
 +\mbox{i}\,\frac{1}{2} Q_\mu \Bigr ) \epsilon_A
\label{covloka}
\end{equation}
is covariant both with respect to the Lorentz and with
respect to the K\"ahler transformations. Indeed it also contains
the K\"ahler connection:
\begin{equation}
Q_\mu  \equiv - \mbox{i}\, \frac{1}{2} \left(\partial_i {\cal
K}\partial_\mu z^i - \partial_{i^\star} {\cal
K}\partial_\mu {\bar z}^{i^\star} \right)
\label{kacon}
\end{equation}
As supersymmetry parameter we choose one of the following form:
\begin{eqnarray}
\epsilon_A &=& e^{f(r)} \xi_a \quad \mbox{$ \chi $ = constant and}\nonumber\\
           && \gamma_0 \chi ^A = \pm \,\mbox{i} \, \varepsilon^{AB} \, \chi _B
           \label{susyparam}
\end{eqnarray}
Using the explicit form of the spin connection for the metric
\eqn{umetric}:
\begin{eqnarray}
\omega^{0a} &=& -\, \partial_a U \, dt \, e^{2U}\nonumber\\
\omega^{ab} &=& 2 \, \partial^a U \, dx^b
\label{spincon}
\end{eqnarray}
and inserting the SUSY parameter \eqn{susyparam} into the
gravitino variation \eqn{gravirule}, from the invariance
condition $ \delta \psi _{A \vert \mu} = 0$ we obtain two equations
corresponding respectively to the case $\mu = 0 $ and to the case
$\mu = a $. Explicitly we get:
\begin{eqnarray}
\frac{dU}{dr} & = & \mp \, 2 \,\mbox{i}\,
\left(\mbox{Im}{\cal  N}\right)_{\Lambda\Sigma}
\, L^\Lambda t^\Sigma \, \frac{e^U}{r^2}
\label{Uequa}\\
\frac{df}{dr} &= &-  \frac{1}{2} \, \frac{dU}{dr}  +  \mbox{i}\frac{1}{2} \,  \Bigl (
\partial_i {\cal K}\frac{d z^i}{dr} -
\partial_{i^\star} {\cal K}\frac{d{\bar z}^{i^\star}}{dr}
\Bigr )\label{fequa}
\end{eqnarray}
 On the other hand setting to zero the gaugino transformation rule
 \eqn{gaugirule} with the SUSY parameter \eqn{susyparam} we obtain:
\begin{equation}
\frac{dz^i}{dr}=\mp 2 \, \mbox{i}\, g^{ij^\star}{\bar f}^\Lambda_{j^\star}
\left( \mbox{Im}{\cal N}\right)_{\Lambda\Sigma} \, \frac{t^\Sigma}{r^2} \, e^U
\label{zequa}
\end{equation}
In obtaining these results, crucial use was made of eq.\eqn{econtr}.
\par
In this way we have reduced the equations for the extremal
BPS saturated black--holes to a pair of first order differential
equations for the metric scale factor $U(r)$ and for the
scalar fields $z^i(r)$. To obtain explicit solutions one should
specify the special K\"ahler manifold one is working with, namely
the specific Lagrangian model. There are, however, some very
general and interesting conclusions that can be drawn in a
model--independent way. They are just consequences of the fact that
the black--hole equations are {\it first order differential equations}.
Because of that there are fixed points
(see the papers \cite{ferkal2,ferkal4,strom3}) namely values either of the metric
or of the scalar fields which, once attained in the evolution
parameter $r$ (= the radial distance ) persist indefinitely. The
fixed point values are just the zeros of the right hand side in
either of the coupled eq.s \eqn{Uequa} and \eqn{zequa}. The fixed
point for the metric equation is $r=\infty $, which corresponds to its
asymptotic flatness. The fixed point for the moduli is $r=0$.  So, independently
from the initial data at  $r=\infty$ that determine the details
of the evolution,  the scalar fields flow into
their fixed point values at $r=0$, which, as I will show,
turns out to be a horizon. Indeed in the vicinity of $r=0$ also the
metric takes a universal form.
\par
Let us see this more closely.
\par
To begin with we consider the equations determining the fixed point
values for the moduli and the universal form attained by the metric
at the moduli fixed point:
\begin{eqnarray}
0 &=& -g^{ij^\star} \, {\bar f}^\Gamma_{j^\star}
\left( \mbox{Im}{\cal N}\right)_{\Gamma\Lambda} \, F^{\Lambda -}_{\mu \nu}
\label{zequato} \\
\frac{dU}{dr} & = & \mp \, 2 \,\mbox{i}\,
\left(\mbox{Im}{\cal  N}\right)_{\Lambda\Sigma}
\, L^\Lambda q^\Sigma \, \frac{e^U}{r^2}
\label{Uequato}
\end{eqnarray}
Multiplying eq.\eqn{zequato} by $f^\Sigma_i$, using the local
special geometry counterpart of eq.\eqn{rigident}:
\begin{equation}
 f^\Sigma_i \, g^{ij^\star} \, {\bar f}^\Gamma_{j^\star}= -\frac{1}{2}\,
 \left(\mbox{Im}{\cal N}\right)^{-1\vert \Sigma\Gamma } -
 {\bar L}^\Sigma \, L^\Gamma
 \label{locident}
\end{equation}
and the definition \eqn{gravifot} of the graviphoton field strength
we obtain:
\begin{equation}
0 = -\frac{1}{2} \, F^{\Lambda -}_{\mu \nu} +\mbox{i}\frac{1}{2} \,
{\bar L}^\Lambda \, T^{-}_{\mu \nu}
\label{ideft}
\end{equation}
Hence, using the definition of the central charge
\eqn{zdef} and eq.\eqn{gaugansz} we conclude that
at the fixed point the following condition is true:
\begin{equation}
0=-\frac{1}{2} \, \frac{t^\Lambda}{4\pi} \, -\, \frac{Z_{fix} \,
{\bar L}_{fix}^\Lambda}{8\pi}
\label{passetto}
\end{equation}
In terms of the previously defined electric and magnetic charges
eq.\eqn{passetto} can be rewritten as:
\begin{eqnarray}
p^\Lambda & = & \mbox{i}\left( Z_{fix}\,{\bar L}^\Lambda_{fix}
- {\bar Z}_{fix}\,L^\Lambda_{fix} \right)\\
q_\Sigma & = & \mbox{i}\left( Z_{fix}\,{\bar M}_\Sigma^{fix}
- {\bar Z}_{fix}\,M_\Sigma^{fix} \right)\\
Z_{fix} &=& M_\Sigma^{fix} \, p^\Lambda\, - L^\Lambda_{fix} \, q_\Lambda
\label{minima}
\end{eqnarray}
which can be regarded as algebraic equations determining the value
of the scalar fields at the fixed point as functions of the electric
and magnetic charges $p^\Lambda, q_\Sigma$:
\begin{equation}
L_{fix}^\Lambda = L^\Lambda(p,q) \, \longrightarrow \,
Z_{fix}=Z(p,q)=\mbox{const}
\end{equation}
\par
In the vicinity of the fixed point the differential equation for the
metric becomes:
\begin{equation}
\pm \, \frac{dU}{dr}=\frac{Z(p,q)}{4\pi \, r^2} \, e^{U(r)}
\end{equation}
which has the approximate solution:
\begin{equation}
\exp[U(r)]\, {\stackrel{r \to 0}{\longrightarrow}}\, \mbox{const} + \frac{Z(p,q)}{4\pi \, r}
\label{approxima}
\end{equation}
Hence, near $r=0$   the metric \eqn{umetric}
becomes of the Bertotti Robinson type:
\begin{eqnarray}
ds^2_{BR} &=& \frac{r^2}{m_{BR}^2}\, dt^2 \, - \, \frac{m_{BR}^2}{r^2} \,
dr^2 \, \nonumber\\
&&- \, m_{BR}^2 \, \left(Sin^2\theta \, d\phi^2 + d\theta^2 \right)
\label{bertrob}
\end{eqnarray}
with Bertotti Robinson mass given by:
\begin{equation}
m_{BR}^2 = \vert \frac{Z(p,q)}{4\pi} \vert^2
\label{brmass}
\end{equation}
In the metric \eqn{bertrob} the surface $r=0$ is light--like and
corresponds to a horizon since it is the locus where the
Killing vector generating time translations $\frac{\partial}{\partial t} $,
which is time--like at spatial infinity $r=\infty$, becomes
light--like. The horizon $r=0$ has a finite area given by:
\begin{equation}
\mbox{Area}_H = \int_{r=0} \, \sqrt{g_{\theta\theta}\,g_{\phi\phi}}
\,d\theta \,d\phi \, = \,  4\pi \, m_{BR}^2
\label{horiz}
\end{equation}
Hence, independently from the details of the considered model,
the BPS saturated black--holes in an N=2 theory have a
Bekenstein--Hawking entropy given by the following horizon area:
\begin{equation}
 \mbox{Area}_H = \, \frac{1}{4\pi} \, \vert Z(p,q) \vert^2
 \label{ariafresca}
\end{equation}
the value of the central charge being determined by eq.s
\eqn{minima}. Such equations can also be seen as the variational
equations for the minimization of the horizon area as
given by \eqn{ariafresca}, if the central charge is regarded
as a function of both the scalar fields and the charges:
\begin{eqnarray}
 \mbox{Area}_H (z,{\bar z})&=& \, \frac{1}{4\pi} \, \vert Z(z,{\bar z},p,q) \vert^2
 \nonumber\\
 \frac{\delta \mbox{Area}_H }{\delta z}&=&0 \, \longrightarrow \,  z = z_{fix}
\end{eqnarray}
\section{
The $p$--branes of string  and $M$--theory and solvable Lie
algebras}
 The solvable Lie algebra structure provides a canonical
parametrization of the scalar field manifold where the fields
associated with the Cartan generators are the {\it generalized
dilatons} which appear in the lagrangian in an exponential way, while
the fields associated with the nilpotent generators
appear in the lagrangian only through polynomials of degree bounded
from above.
\par
Since the fermion transformation rules and the associated central charges
of all maximally extended supergravities are expressed solely in
terms of the coset representative $L\left(\phi\right)$ (see \cite{voialtri}),
the  method for the derivation of
extremal  solutions, which in the previous section was applied to the
case of $D=4, N=2$ black--holes, can now be extended to the case of
extremal $p$--brane solutions in $D=10-r$. The canonical parametrization
of the scalars through solvable Lie algebras hints to a complete
solubility of the corresponding first order equations, namely of the
analogues of eq.s\eqn{Uequa},\eqn{fequa}. This investigation is work
in progress \cite{progre} by the author and the same collaborators as in
\cite{solvab1}, \cite{solvab2}.
\par
To illustrate the idea we just recall the results obtained in the
literature for $p$--brane solutions. In \cite{stellebrane} the
following bosonic action was considered:
\begin{eqnarray}
{S^\prime}_{D}&=& \int \, d^Dx \, {\cal L}_D \nonumber \\
 {{\cal L}^\prime}_D   &=& \sqrt{\mbox{det}g} \Bigl (  -2 \, R[g]\, -  \,
  \frac{1}{2} \partial^\mu \phi \, \partial_\mu \phi
   -\frac{(-1)^{n-1}}{2{n!}} \, \exp[-a\phi] \, F_{\mu_1 \dots
  \mu_n}^2 \Bigl )
  \label{paction}
\end{eqnarray}
where $F_{\mu_1 \dots \mu_n}$ is the field strength of an $n-1$--form
gauge potential, $\phi$ is a dilaton and $a$ is some real number. For various values of
$n$ and $a$,  ${S^\prime}_{D}$ is a consistent truncation of some
(maximal or non maximal) supergravity bosonic action $S_D$
 in dimension $D$. By consistent truncation we mean that a
subset of the bosonic fields have been put equal to zero but
in such a way that all solutions of the truncated action are also
solutions of the complete one. The fields that have been
deleted are:
\begin{enumerate}
\item {all the nilpotent scalars}
\item {all the Cartan scalars except that which appears in front of the
$F_{\mu_1 \dots \mu_n}$ kinetic term.}
\item {all the other gauge $q$--form potentials except the chosen
one}
\end{enumerate}
For instance if we choose:
\begin{equation}
a=1 \quad \quad n=\cases{3 \cr 7\cr}
\label{het}
\end{equation}
eq.\eqn{paction} corresponds to the bosonic low energy action of $D=10$
heterotic superstring (N=1, supergravity) where the $E_8\times
E_8$ gauge fields have been deleted. The two choices $3$ or $7$ in
eq.\eqn{het} correspond to the two formulations (electric/magnetic)
of the theory. Other choices correspond to truncations of the type IIA or
type IIB action in the various intermediate dimensions $4\le D\le
10$. Since the $n-1$--form $A_{\mu_1\dots\mu_{n-1}}$ couples to the world volume
of an extended object of dimension:
\begin{equation}
p = n-2
\label{interpre}
\end{equation}
namely a $p$--brane, the choice of the truncated action \eqn{paction}
is motivated by the search for $p$--brane solutions of supergravity.
According with the interpretation \eqn{interpre} we set:
\begin{equation}
  n=p+2  \qquad  d=p+1 \qquad
{\tilde d}= D-p-3
\label{wvol}
\end{equation}
where $d$ is the world--volume dimension of the electrically charged
{\it elementary} $p$--brane solution, while ${\tilde d}$ is
the world--volume dimension of a magnetically charged {\it solitonic}
${\tilde p}$--brane with ${\tilde p} = D-p-4$. The distinction between
elementary and solitonic is the following. In the elementary case
the field configuration we shall discuss is a true vacuum solution of
the field equations following from the action \eqn{paction} everywhere in
$D$--dimensional space--time except for a singular locus of dimension
$d$. This locus can be interpreted as the location of an elementary $p$--brane
source that is coupled to supergravity via an electric charge   spread
over its own world volume. In the solitonic case, the field
configuration we shall consider is instead a bona--fide solution of
the supergravity field equations everywhere in space--time without
the need to postulate external elementary sources. The field energy
is however concentrated around a locus of dimension ${\tilde p}$.
Defining:
\begin{equation}
\Delta = a^2 +2\, \frac{d {\tilde d} }{ D-2}
\end{equation}
it was shown in \cite{stellebrane} that action \eqn{paction} admits
the following elementary $p$--brane solution
\begin{eqnarray}
ds^2 & =& \left(1+\frac{k}{ r^{\tilde d} } \right)^{-  \frac {4\, { \tilde d}} {\Delta (D-2)}}
\, dx^\mu \otimes dx^\nu \, \eta_{\mu\nu}
- \left(1+\frac{k}{ r^{\tilde d} } \right)^{ \frac {4\, {d}} {\Delta (D-2)}}
\, dy^m \otimes dy^n \, \delta_{mn}\nonumber \\
F &= &\lambda (-)^{p+1}\epsilon_{\mu_1\dots\mu_{p+1}} dx^{\mu_1}
\wedge \dots \wedge dx^{\mu_{p+1}}
\wedge \, \frac{y^m \, dy^m}{r} \, \left(1+\frac{k}{r^{\tilde
d}}\right )^{-2} \, \frac{1}{r^{\tilde d}}\nonumber\\
e^{\phi(r)} &=& \left(1+\frac{k}{r^{\tilde d}}\right)^{-\frac {2a}{\Delta}}
\label{elem}
\end{eqnarray}
where $x^\mu$, $(\mu=0,\dots ,p)$ are the coordinates on the $p$--brane world--volume,
$y^m$, $(m=D-d+1,\dots ,D)$ are the transverse coordinates, $r \equiv \sqrt{y^m y_m}$,
$k$ is the value of the electric charge and:
\begin{equation}
\lambda= 2\, \frac{{\tilde d} \, k}{\sqrt{\Delta}}
\end{equation}
The same authors show that that action \eqn{paction} admits also
the following solitonic ${\tilde p}$--brane solution:
\begin{eqnarray}
ds^2 & =& \left(1+\frac{k}{ r^{d} } \right)^{- \frac {4\, {   d}} {\Delta (D-2)}}
\, dx^\mu \otimes dx^\nu \, \eta_{\mu\nu}
- \left(1+\frac{k}{ r^{ d} } \right)^{ \frac {4\, {\tilde d}} {\Delta (D-2)}}
\, dy^m \otimes dy^n \, \delta_{mn}\nonumber \\
{\tilde F} &= &\lambda  \epsilon_{\mu_1\dots\mu_{{\tilde d}}p} dx^{\mu_1}
\wedge \dots \wedge dx^{\mu_{\tilde d}}\wedge \frac{y^p}{r^{d+2}} \nonumber\\
e^{\phi(r)} &=& \left(1+\frac{k}{r^{d}}\right)^{\frac {2a}{ \Delta}}
\label{solit}
\end{eqnarray}
where the $D-p-2$--form ${\tilde F}$ is the dual of $F$, $k$ is now the magnetic charge
and:
\begin{equation}
\lambda= - 2\, \frac{{\tilde d} \, k}{\sqrt{\Delta}}
\end{equation}
These  $p$--brane configurations are solutions of the second order
field equations obtained by varying the action \eqn{paction}.
However, when \eqn{paction} is the truncation of
a supergravity action both \eqn{elem} and \eqn{solit} are also the
solutions of a {\it first order differential system of equations}.
This happens because they are BPS--extremal $p$--branes which
preserve a fraction of the original supersymmetries. For instance
consider the $10$--dimensional case where:
\begin{equation}
 D=10 \qquad d=2 \qquad {\tilde d}= 6 \qquad
a=1 \qquad \Delta = 4 \qquad \lambda = \pm 6 k
\label{values}
\end{equation}
so that the elementary string solution reduces to:
\begin{eqnarray}
ds^2 &=& \exp[2 U(r)] \, dx^\mu \otimes dx^\nu -
\exp[-\frac{2 }{3}  U(r)] \, dy^m \otimes dy^m
 \nonumber\\
\exp[2 U(r)]&=& \left(1+\frac{k}{r^6}\right)^{-3/4}\nonumber \\
F &=& 6k \, \epsilon_{\mu\nu} dx^\mu \wedge dx^\nu \wedge \frac{y^m dy^m}{r} \,
\left( 1+\frac{k}{r^6}\right)^{-2} \, \frac{1}{r^7}\nonumber\\
\exp[\phi(r)] &=&  \left(1+\frac{k}{r^6}\right)^{-1/2}
\label{stringsol3}
\end{eqnarray}
As already pointed out, with the values \eqn{values}, the action \eqn{paction}
is just the truncation of heterotic supergravity where, besides the
fermions, also the $E_8\times E_8$ gauge fields have been set to zero.
In this theory the SUSY rules we have to consider are those of the
gravitino and of the dilatino.
They read:
\begin{eqnarray}
\delta \psi _\mu &=& \nabla_\mu \epsilon\, +\, \frac{1}{96} \,
\exp[\frac{1}{2} \phi] \, \Bigl ( \Gamma_{\lambda\rho\sigma\mu}
+\, 9\,  \Gamma_{\lambda\rho} \, g_{\sigma\mu} \Bigr ) \, F^{\lambda\rho\sigma}
\, \epsilon \nonumber\\
\delta \chi &=& \mbox{i}\, \frac{\sqrt{2}}{4} \, \partial^\mu \phi \,
\Gamma_\mu \epsilon  -\,  \mbox{i}\, \frac{\sqrt{2}}{24} \, \exp[-\frac{1}{2}\phi ] \,
\Gamma_{\mu\nu\rho} \, \epsilon \, F^{\mu\nu\rho}
\label{susvaria}
\end{eqnarray}
Expressing the $10$-dimensional gamma matrices as tensor products of
the $2$--dimensional gamma--matrices $\gamma_\mu$ ($\mu=0,1$) on the
$1$--brane world sheet with the $8$--dimensional gamma--matrices
$\Sigma_m$ ($m=2,\dots, 9$) on the transverse space it is easy to check
that in the background \eqn{stringsol3}
the SUSY variations
\eqn{susvaria} vanish for the following choice of the parameter:
\begin{equation}
\epsilon   =   \left( 1+\frac{k}{r^6}\right)^{-3/16} \, \epsilon_0
\otimes \eta_0
\label{carmenpara}
\end{equation}
where the constant spinors $\epsilon_0$ and $\eta_0$ are respectively
$2$--component and $16$--component and have both positive chirality:
\begin{equation}
\matrix{ \gamma_3 \, \epsilon_0 = \epsilon_0 &  \Sigma_0 \, \eta_0 = \eta_0}
\label{chiralcondo}
\end{equation}
Eq.\eqn{chiralcondo} is the $D=10$ analogue of eq.\eqn{parcondicio}.
Hence we conclude that the extremal $p$--brane solutions of all
maximal (and non maximal) supergravities can be
obtained by imposing the supersymmetry invariance of the background
with respect to a projected SUSY parameter of the type
\eqn{carmenpara}. \par
In the maximal case a general analysis of the resulting evolution equation
for the scalar fields in the solvable Lie algebra representation is
work in progress \cite{progre}.

%%%%%%%%%%%%%%%%%%%%%%%%%%%%%%%%%%%%%%%%%%%%%%%%%%
% End of File action.tex %%%%%%%%%%%%%%%%%%%%%%%%%%%
%%%%%%%%%%%%%%%%%%%%%%%%%%%%%%%%%%%%%%%%%%%%%%%%%%%%

\end{document}